\documentclass[12pt,a4paper]{article}
\usepackage[margin=1in]{geometry}
\usepackage[utf8]{inputenc}
\usepackage{natbib}
\usepackage{hyperref}
\usepackage{verbatim}
\usepackage{amsmath}
\usepackage{amsfonts}
\usepackage{amssymb}
\usepackage{amsthm}
\usepackage{mathtools}
\usepackage{bm}
\usepackage{xparse}
\usepackage{graphicx}
\usepackage{xcolor}
\usepackage{textcase}
\usepackage{url}
\usepackage{lipsum}
\usepackage{appendix}
\usepackage{epstopdf}
\usepackage{setspace}
\usepackage{centernot}
\usepackage{braket}

\usepackage{titlesec}

\usepackage[pagewise]{lineno}

\usepackage{float}

\usepackage{caption}
\usepackage{subcaption}

\titleformat{\section}
  {\normalfont\scshape \filcenter \large}{\thesection}{1em}{}

\titleformat{\subsection}
  {\normalfont\scshape \filcenter}{\thesection}{1em}{}

\usepackage{indentfirst}

\usepackage{amsthm}

\usepackage{authblk}

\usepackage{dsfont}

\begin{document} 

\title{Locality, Probability and the Relevance of History}

\author{Sebastian Horvat\\
University of Vienna, Department of Philosophy, Universitätsstraße 7, 1010 Vienna, Austria}
%\affil{University of Vienna, Department of Philosophy, Universitätsstraße 7, 1010 Vienna, Austria}
\date{}                     
\setcounter{Maxaffil}{0}
\renewcommand\Affilfont{\itshape\small}

%\date{\today}

\maketitle

\begin{abstract}
Relativistic theory and experiments teach us that causal influences are constrained by the speed of light. Deterministic theories incorporate this lesson in \textit{functional} terms: whether an event occurs or not is functionally determined by events that occur in its immediate causal past. In this paper we suggest a natural generalization of this constraint that applies to probabilistic theories: the \textit{probability} that an event occurs is to be functionally determined by the \textit{probability} of the event-occurrences in its immediate causal past. This constraint differs from Bell’s criterion of local causality and turns out to imply it only under an additional assumption that requires history to be, in a sense, \textit{ir}relevant. Bell-inequality-violating quantum phenomena are thus arguably \textit{not} in tension with relativity, locality, ``statistical independence'', nor with a supposed ``realism''. What they instead show is that history is, in a sense, \textit{relevant}.

\end{abstract}

\section*{Introduction}

John S. Bell's criterion of \textit{local causality}, conjoined with relatively innocuous assumptions, is incompatible with certain empirically supported predictions of quantum theory.\footnote{See Myrwold et al. (2024) for an overview of the ``relatively innocuous'' assumptions that figure in Bell's theorem, such as the assumption of the statistical independence of the properties of what is measured from the methods by which it is measured, and the supposed assumption that quantum measurements have unique outcomes. For a detailed analysis of the criterion of local causality, consult Norsen (2013).} This incompatibility is puzzling, in that, on the one hand, the said assumptions truly \textit{are} innocuous, and on the other, the said criterion appears to be backed by the empirically supported theory of relativity. In this paper we will argue that it is the second horn of the latter dilemma that is the faulty one: relativistic theory and experiments do not support Bell's criterion, but \textit{another} criterion, whereby the new criterion implies the old one only if an \textit{additional} assumption holds - the ``irrelevance of history'' - an assumption whose negation is intelligible. The empirical violation of Bell's inequalities is thus arguably not in tension with relativity. It rather shows that the said additional assumption is false. 

Before diving head-front into a detailed exposition of our new criterion, let us offer some informal remarks that are to indicate where the fault in Bell's original criterion lies. 

\section*{I. On the Criterion of Local Causality}

\begin{figure}[H]
\centering
\includegraphics[width=0.8\linewidth]{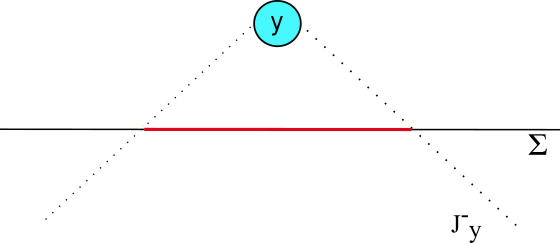}
\caption{Relativity teaches us that what happens at $y$ is determined by what happens at the red-colored intersection of its causal past $J^-_y$ and an arbitrarily close past Cauchy surface $\Sigma$.}

\label{fig1}
\end{figure}

The relativistic lesson that the original criterion of local causality is intended to explicate is that the causes of any event-occurrence lie arbitrarily close to it in its past lightcone: for example, my writing of these words here-and-now is presumably caused by what has happened in my immediate causal past, and not, say, by whatever might be going on in distant galaxies.\footnote{It is to be noted that both our presentation of the \textit{motivation} for the criterion of local causality and our presentation of its \textit{formulation} do not \textit{exactly }coincide with Bell's own presentation (e.g. in his (2004, p. 232-248)), but are rather to be understood as something akin to a ``rational reconstruction'' thereof.} Eliminating the somewhat problematic notion of ``causality'' in favor of ``determination'', the lesson reads, more precisely, as follows: what occurs, or is the case, in a spatiotemporal region $y$, is determined by what occurred, or was the case, at the intersection of the past lightcone $J^{-}_y$ of $y$ and a past Cauchy surface $\Sigma$ that is arbitrarily close to $y$ - see Figure \ref{fig1}. 

The above constraint is paradigmatically incorporated by source-free classical electrodynamics, which dictates that the electric and magnetic field values in some region $y$ are to be \textit{functionally determined} by the respective field values at the intersection of $J^{-}_y$ and an arbitrarily close past Cauchy surface $\Sigma$. In other words, while the theory allows many ways for the fields to be spread across space and time, these ways are constrained in such a way, that any two possible field configurations that agree on the field values at some $\Sigma \cap J^{-}_y$, must also agree on the field values at $y$. The ``ways'' referred to here correspond to the solutions of the theory's equations of motion, and can be thought of as the phenomena or worlds that the theory regards to be physically possible, the ways the world could be according to the theory. Our relativistic lesson, formulated in terms of functional determination, thus does not constrain \textit{individual }ways-worlds-phenomena that the theory deems to be possible; it rather constrains \textit{relations} between pairs of possible ways-worlds-phenomena, so that any pair that agrees on something - on what is the case in the immediate causal past of $y$ - must also agree on something else - on what is the case at $y$.

Bell introduced his criterion of local causality in order to rigorously distinguish those physical theories that incorporate our relativistic lesson from those that do not: source-free classical electrodynamics and other classical relativistic field theories are to be exemplarily deemed as locally causal, whereas theories such as Newtonian mechanics are to be excluded from this lot. The general criterion however could not, and ought not to, be formulated exactly as above, in terms of functional determination, since no \textit{in}deterministic theory can possibly satisfy such a constraint: indeed, according to an indeterministic theory, what happens here-and-now is not determined by what happened in the past, let alone by what happened in the \textit{immediate causal} past. This is problematic in so far as the phenomena studied by quantum theory seem not to admit of a deterministic account, and yet \textit{do} bear a shadow of our relativistic lesson, in that they do not seem to allow for a superluminal transfer of information (see e.g. Ghirardi et al., 1980). The criterion therefore needs to be relaxed, if it is to have a chance at identifying those theories that can possibly account for quantum phenomena, while complying with relativistic theory and experiments.

\begin{figure}[H]
\centering
\includegraphics[width=0.8\linewidth]{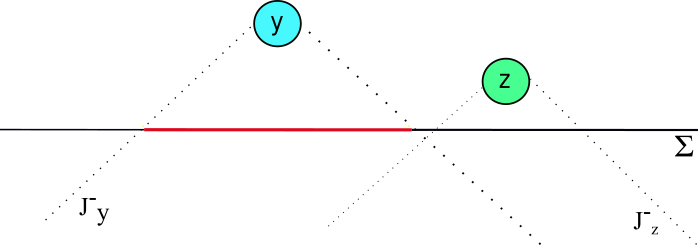}
\caption{According to a locally causal theory, the statistical correlation between events possibly occurring at $y$ and events possibly occurring at $z$ is screened off by events occurring at the red-colored $\Sigma \cap J^{-}_y$. Note that the past lightcone of $z$ does \textit{not} intersect $\Sigma \cap J^{-}_y$. Indeed, if it \textit{did} intersect it, then there would be no relativistically backed reason to expect the just cited constraint to hold.}

\label{fig2}
\end{figure}

The way Bell chose to relax the deterministic criterion of local-causality-as-functional-determination is by changing the object that is supposed to be determined according to a locally causal indeterministic theory: it is not what occurs, or is the case, but only the \textit{probability} thereof, that is supposed to be determined by what occurred, or was the case, in the immediate past lightcone. More precisely, let $y$ be an arbitrary spatiotemporal point, $\Sigma \cap J^{-}_y$ an arbitrary slice of its past lightcone and $z$ an arbitrary point that is spacelike separated from $y$, and whose past lightcone does not intersect $\Sigma \cap J^{-}_y$, as in Figure \ref{fig2}. According to a locally causal indeterministic theory, any physically possible phenomenon $\mathcal{P}$ is such that its exhibited probability $P(a|\lambda b)$ that some event $a$ occurs at $y$, conditioned on events $\lambda$ having occurred at $\Sigma \cap J^{-}_y$ and event $b$ having occurred at $z$, satisfies the following relation

\begin{equation}
    P(a|\lambda b) = P(a|\lambda).
\end{equation}

In other words, according to a locally-causal indeterministic theory, the probability of any event-occurrence in any phenomenon is fixed once conditioned on sufficiently many event-occurrences situated in its immediate past lightcone. As it is well known, applying this criterion to EPR-like experiments, and conjoining it with certain hardly dubitable assumptions - such as the statistical independence of the state of what is measured from the choice of how it is to be measured\footnote{For a recent discussion of this assumption, see Waegell $\&$ McQueen (2025). For a critical discussion of the related issue of ``superdeterminism'', see Baas \& Le Bihan (2023).} - leads to the validity of Bell's inequalities, and thus to a conflict with their empirical \textit{non}-validity.

The probabilistic criterion sketched above is supposed to be a probabilistic generalization of the sense of determination present in the deterministic criterion: events in the past lightcone of $y$ are not required to determine which particular event occurs at $y$, but are only required to determine the probability distribution of events that may occur at $y$. It is however unclear in what exact sense the current criterion can be seen as a \textit{generalization} of the deterministic criterion, due to there being an important disanalogy between the two. As noted beforehand, the deterministic criterion of local causality was articulated in terms of functional determination, and thus appealed, explicitly or implicitly, to a space of possible ways-worlds-phenomena posited by the theory under examination. The current probabilistic criterion, on the contrary, does not seem to appeal to any such space of possibilities, as it refers only to \textit{one} probability distribution $P$ that describes \textit{one} phenomenon $\mathcal{P}$. In other words, while the deterministic criterion concerns \textit{relations} between different phenomena - corresponding to different initial or boundary conditions - the probabilistic criterion of local causality concerns separately each \textit{individual }phenomenon - each corresponding to is own initial or boundary condition.

To be sure, the intention behind the probabilistic criterion \textit{is} to refer to a plurality of possible phenomena. The criterion namely intends to say that a theory is locally causal if all pairs of allowed phenomena that agree that $\lambda$ occurs at $\Sigma \cap J^{-}_y$, but which disagree on what occurs at $z$, nevertheless agree on the probability that $a$ occurs at $y$: $P(a|\lambda b) = P(a|\lambda b')$. The issue is however that the probability distribution $P$ referred to here is not explicitly associated to phenomena in which $\lambda$ and $b$, or $\lambda$ and $b'$, occur deterministically; it is instead associated to the original phenomenon $\mathcal{P}$, which may generally figure unsharp probability distributions of events that occur at the respective locations $\Sigma \cap J^{-}_y$ and $z$. Making explicit the desired reference to spaces of possible ways-worlds-phenomena accordingly requires more care, especially in the evaluation of the physical significance of the just cited conditional probabilities that enter Bell's criterion. 
%which may at the moment admittedly appear to be a mere push for pedantry.

%This observation opens up a logical gap: even though the original phenomenon $\mathcal{P}$ may be described by distribution $P$, this does not imply that there is another phenomenon $\mathcal{P}_{\lambda,b}$ in which $\lambda$ and $b$ occur deterministically, and which are described by this same duly conditionalized distribution $P(\cdot|\lambda b)$. Stated succinctly, it is logically possible that the conditional probabilities referred to in the probabilistic criterion do not present adequate descriptions of phenomena in which $\lambda$ and $b$, or $\lambda$ and $b'$ are fixed. If that is so, then a criterion formulated explicitly in terms of pairs of distributions $P_{\lambda, b}$ and $P_{\lambda, b'}$ - that describe pairs of phenomena $\mathcal{P}_{\lambda,b}$ and $\mathcal{P}_{\lambda,b'}$ - has a chance at substantially differing from the original criterion, it being formulated in terms of conditional probabilities $P(\cdot|\lambda b)$ and $P(\cdot|\lambda b')$ generated from one distribution $P$, which describes one phenomenon $\mathcal{P}$.

Besides not being formulated in terms of spaces of possible ways-worlds-phenomena, there is a \textit{further} disanalogy that separates the deterministic and the probabilistic criteria. The deterministic criterion namely spoke of the determination of certain events on behalf of other events. The relata that figured therein were accordingly of the same kind, namely event-occurrences. The probabilistic criterion however breaks the symmetry between the relata of determination, in that event-occurrences are not supposed to determine other event-occurrences, but are instead supposed to determine \textit{probability distributions} of event-occurrences. But why should we impose that event-occurrences themselves determine probability distributions of other event-occurrences? Should we not rather impose that the \textit{probability distributions} of some event-occurrences determine the probability distributions of other event-occurrences, thereby retaining the symmetry in kind between the relata of determination? While probability distributions surely cannot stand in \textit{causal} relations to each other - arguably only events or objects can do so - why should they not be capable of standing in relations of \textit{determination} to each other? Indeed, if we are ready - as per Bell's criterion - to regard probability distributions to be possibly determined by event-occurrences, should we not also be ready to regard them as possibly being the ones that do the determining?

The above two disanalogies point at the logical possibility of a criterion that might differ from Bell's probabilistic criterion of local causality, one that (i) explicitly adheres to spaces of possible ways-worlds-phenomena, and that (ii) is formulated in terms of a relation of determination that holds among probability distributions. To be sure, point (i) may at the moment appear to be a mere pedantic explication of something that is already present in the original criterion; also, regarding point (ii), it may currently seem unintelligible for probability distributions themselves to be the ones that do the determining, since conditional probabilities in physics are normally defined by conditioning on events, and not on probability distributions themselves. It is however our intention, in what follows, to articulate this possibility as a real and intelligible one, to show that the criterion that it presents \textit{does} substantially differ from Bell's, to maintain that \textit{this} is the criterion that is actually supported by relativistic theories and experiments, and to consequently, and tentatively, propose what the empirical violation of Bell's inequalities actually implies.

\section*{II. Events, Worlds, World-Theories}

In the previous section we briefly and informally noted two reasons for which Bell's criterion of local causality is to be regarded as possibly not directly supported by relativistic theory and experiments. In this section our goal is instead constructive: it is to cautiously argue our way towards the criterion, or ``principle'', that \textit{is} actually supported by relativity. We will thereby recapture the remarks from the previous section, albeit in a clearer form.

Note first that relativistic theory and experiments certainly do not support only one single principle, but a plurality thereof, and that some of these may refer to \textit{objects} or \textit{systems} and their mutual interactions. For example, we have all heard of the principle that mandates that a material object cannot travel faster than light, or of the one that prohibits that the state of a system be instantaneously changed due to the change of the state of a distant system.\footnote{Relativity also supports certain technological and information-theoretic principles that constrain the capabilities of agents like us: for example, nobody is presumably capable of inducing an instantaneous change in a distant system; or, no community is able to communicate superluminally. In this paper we are however not going to say anything about these, as we are only focusing on \textit{physical} principles, that is, on those principles that concern what is or can \textit{be the case}, rather than what can \textit{be done}.} While some such principle, phrased in terms of temporally persisting objects-systems, might be valid in many circumstances - in those circumstances where persisting and relatively well localized objects-systems can be identified in the first place - we will here focus on a possible formulation that refers only to \textit{events}, without reference to underlying objects that may partake in these events - `events' here meaning the type of things that can \textit{occur} or \textit{be the case} in some spatiotemporal region. 

We will not delve here into a detailed analysis of the relationship between principles that concern objects and principles that concern events. We will instead only briefly note that any principle that speaks of the change of an object - e.g. of the change of a system's state, or the movement of a material body - is bound to also imply something about events. The reason for this is simply that a changing object changes in time, thus allowing for any such change to be described in terms of a sequence of events. The converse is however not necessarily the case, in that a sequence of event-occurrences does not strictly imply the presence of an underlying subsisting object, whose changes are presumably to be manifested in form of the said events. There thus \textit{seems} to be a sense in which principles that concern events are more fundamental than any dynamical principle that concerns the change of objects.\footnote{\textit{Non}-dynamical principles that concern objects might on the other hand as well imply nothing about event-occurrences: e.g. that for each particle-type there is its corresponding antiparticle-type does not seem to imply any event-principle.} We will however not attempt to turn this intuition into a rigorous argument here, but just regard it as an initial stimulus for a focus on event-principles.

%However that may be, Bell's own criterion was phrased in terms of what we are calling events, and what he was calling beables: things that are the case in some spatiotemporal region
Some relativistic event-principles have of course already been proposed; indeed, Bell's own criterion of local causality amounts to one, in that the ``beables'' that he was talking about are what we are referring to as ``events''.\footnote{Other event-principles includes cognates of Bell's original criterion, such as various versions of ``stochastic Einstein locality'' (Hellman, 1982; Butterfield, 2007; see also Henson, 2005).} These principles - including the one that we will later suggest - can be understood as tentative precisifications-explications of the aforementioned pre-precise relativistic lesson, which, to repeat, reads as follows: whether an event occurs or not depends entirely on what occurred on, and within, its immediate past lightcone - that is, in its immediate ``causal past''. In other words, an event-occurrence is presumably entirely ``determined'' by what occurs in its immediate causal past, and unaffected by what occurs at its spacelike separation. 

%Before starting with our precisification, let us, already at this stage, roughly differentiate two different pre-precise principles that are to be turned into two different precisified principles: (i) the principle of \textit{relativistic causality} is to mandate that an event-occurrence is determined by what occurs in its causal past, whereas (ii) the principle of \textit{locality} is to mandate more strongly that an event-occurrence is determined by what occurs in its \textit{immediate} causal past, or, so to say, in the infinitesimal causal vicinity of the said event. Principle (i) is expected to exclude ``spatially non-local'' interactions between spatially separated objects, whereas principle (ii) is in addition expected to rule out ``temporally non-local'' interactions between temporally separated objects.

The pre-precise principle speaks of dependencies or determinations holding between event-occurrences. Now, that an event-occurrence $e$ depends on - or is determined by - an event-occurrence $f$, can arguably not be explicated merely by reference to \textit{one} phenomenon that includes instances of $e$ and $f$. The dependence or determination can instead be articulated by reference to \textit{multiple} possible phenomena, and to the way instances of $e$ and $f$ relate to each other across such phenomena: roughly speaking, that $f$ depends only on $e$ - or that $f$ is determined entirely by $e$ -  amounts to $f$ occurring in all those possible phenomena in which $e$ occurs. Note that the ``possible phenomena'' that we have in mind here are \textit{physical}, rather than merely logico-mathematical: while we can certainly linguistically-mathematically represent various phenomena that do not respect relativistic constraints, empirical evidence indicates that only those representations that do satisfy these constraints have a chance at representing \textit{physically possible} phenomena. 

The articulation of our principle therefore appears to require an appeal to a collection of physically possible phenomena. Now, there are two kinds of collections of physical possibilities that we can appeal to here. We can either appeal to a certain ideal collection of physical possibilities - ``ideal'' in the sense of potentially remaining outside of our cognitive reach - a collection that contains only those phenomena that \textit{really are} physically possible, and that we might as well only have a chance at encountering at the Percean end of inquiry. \textit{Or}, we can appeal to the possibilities posited by a physical theory - not to what \textit{really is} physically possible, but to what a theory \textit{states to be} physically possible - possibilities that \textit{are} within our reach, and that can \textit{at best} aspire to coincide with the formerly mentioned ideal collection. A formulation that would follow the first option would articulate our principle as a tentative truth that concerns physically possible phenomena themselves - with the presupposition that whether a phenomenon is possible or impossible is, in a certain sense, already determined in itself, independently of the theories and other technologies that we might develop in order to grasp these possibilities. On the other hand, a formulation that follows the second option would not appeal to phenomena and their possibilities themselves, but to properties of a cultural object, or in our case, to the physical-modal profile of a theory.

The two approaches are dual to each other, in that, stated schematically, a theory aspires to coincide with what is the case, and, conversely, what is the case is what is articulable by an ideal theory. Here we will however choose the second way, the way of theory, simply in order not to exclude the empiricist from our discussion, the empiricist being the person that treats all metaphysics, and especially \textit{modal} metaphysics, with an element of suspicion. While someone might doubt the intelligibility of thinking that phenomena are, in a certain sense, possible or impossible in themselves, we are all capable of turning our focus on theories and on their properties. We will therefore articulate our principle as a constraint on modal profiles of theories. This is after all the way that was also chosen by Bell, who did not speak of locally causal phenomena, but of locally causal \textit{theories}.

Proceeding with another remark that concerns our pre-precise principle, it is to be noted that it appeals essentially to \textit{past lightcones} of events. Since the past lightcone that can be associated to any real event - such as the one elicited by my current writing - appears to extend itself indefinitely into the past, this motivates us - while certainly not necessitating us - to focus on theories that posit possible phenomena that are, in a sense, extended throughout the whole of space and time, and that therefore contain the causal pasts and futures of all of the events occurring within these phenomena. Such possible phenomena can be understood as possible ways the entire universe could be, in the entirety of its past and of its future. In other words, the theories that we will be interested in are those that induce sets of \textit{physically possible worlds}, as some like to call them. We will accordingly refer to such theories as ``world-theories''.

A world-theory will, roughly speaking, specify a collection of possible worlds, whereby each world will be represented by a specification of spatiotemporally related events that constitute that world. Moreover, each such world is to be equipped with some basic causal structure that specifies, for each event $e$ within that world, the set of all those events that lie in the causal past or future of $e$. We will, for simplicity, consider only those world-theories that posit a fixed spatiotemporal background and that define their various worlds by populating this fixed background with different configurations of events. Our proposal will however not depend in any essential way on the positing of such a background, and could accordingly be extended to theories that articulate the spatiotemporal character of events purely in terms of relations holding between them.

Note again that our pre-precise principle relies essentially only on the notion of the ``past lightcone'' or the associated ``causal past'' of an event, containing no other reference to fine-grained spatiotemporal structure. The spatiotemporal backgrounds relative to which our world-theories will be defined can thus be left quite general and unspecified. More precisely, the spatiotemporal background will be modelled by any Lorentzian manifold $\mathcal{M}$ that allows for the definition of a relation `$<$' of ``strict causal precedence''. That is, points $x,y \in M$ satisfy $x<y$ if and only if there exists a future-directed non-spacelike curve from $x$ to $y$, \textit{and} $x \neq y$. Relation $<$ accordingly induces the definition of the causal past $J^{-}_Y$ of any region $Y \subset M$
\begin{equation}
    J^-_Y \equiv \left\{x \in M| \exists y \in Y:x<y \right\},
\end{equation}
and the analogous definition of its causal future $J^{+}_Y$. For reasons that will be apparent later, it will also come of use to assume that $\mathcal{M}$ is \textit{globally hyperbolic}, thus allowing for a foliation into Cauchy surfaces $\left\{\Sigma_t \right\}_{t \in \mathbb{R}}$: each $\Sigma_t$ contains all those spatiotemporal locations that share the same moment in time $t$, the temporal coordinate being associated to some inertial reference frame associated to the foliation in question. We will henceforth for simplicity refer to Cauchy surfaces simply as `surfaces'. Also, we will for simplicity say that $\Sigma_t<\Sigma_{t'}$ holds if and only if $t<t'$ holds.

In sum, whenever reference is made to a spacetime manifold $\mathcal{M}$, it will be assumed that it contains sufficient structure for the definition of (i) a relation of strict causal precedence $<$, and (ii) a foliation into surfaces $\left\{\Sigma_t \right\}_{t \in \mathbb{R}}$. With this basic spatiotemporal machinery in place, we are now ready to define \textit{world-theories}, that is, theories that posit sets of possible worlds, worlds that are constructed by populating a fixed spatiotemporal background with various configurations of events. 

\bigskip

\textbf{Definition 1.} (World-theories) A \textit{world-theory} is a triple $(\mathcal{M},E,\mathcal{W})$, where $\mathcal{M}$ is a spacetime manifold, $E$ is a set, and $\mathcal{W}$ is a set that contains functions $W: \mathcal{M} \rightarrow E$. 

\bigskip

We will now say how the mathematical structure introduced by the latter definition is intended to represent physically possible worlds. Firstly, set $E$ is intended to represent \textit{mutually exclusive} event-\textit{types}: this means that any two \textit{particular} events that instantiate the event-types referred to by two different elements of $E$ cannot co-occur at the very same spatiotemporal location. For example, two different elements of $E$ may be taken to refer to ``a presence of a red apple'' and to ``a presence of a green apple'' respectively, because one and the same sufficiently small spatiotemporal region cannot host both a green and a red apple. On the other hand, two different elements of $E$ \textit{cannot} be taken to refer to ``a presence of a red apple'' and to ``a presence of a fruit'', since one and the same particular event may instantiate both of the latter event-types. Secondly, and accordingly, each function $W: \mathcal{M} \rightarrow E$ associates, to each spatiotemporal location represented by $x \in \mathcal{M}$, an event-type represented by $W(x)$. A world-theory thereby posits a set of possible worlds represented by elements of $\mathcal{W}$, that is, by different ways in which event-types in $E$ can be distributed across space and time.\footnote{For easier readability we will henceforth refer to event-types simply as `events': for example, when we say that two worlds posit the same event at some location $x$ - or that they agree concerning what occurs at $x$ - we will mean that these worlds associate the same event-\textit{type} to the said location.} 

Let us make a brief and cautionary note on the empirical content of world-theories. Namely, note that these theories need not be \textit{cosmological} theories: their sets of worlds may as well not contain any elements that are plausibly to be regarded as adequate representations of the actual history of our universe. A world-theory might for instance posit a set of worlds, each of which contains events arising out of movements of only \textit{one} electron, worlds that are thus all strictly in disaccord with our actual history, but that we still have some reason to believe to be, in a certain sense, physically possible, had our history been different. The evidence that supports a world-theory can of course not be obtained by running back to some supposed past boundary of our actual universe and reloading the initial conditions. Rather, every relatively well isolated phenomenon that we can point at, study and manipulate \textit{within} our universe can be thought of as hiding a microcosmos within it, as a synecdoche of what our universe could have been: for example, experiments involving isolated single electrons can teach us about what a universe containing a single electron could have been. %Relatedly, the past boundaries posited by world-theories are not to suggest that our actual universe need to have a past boundary, but are merely a reflection of the fact that any physical phenomenon that we can ever encounter within our universe - and that we can tentatively model using the worlds posited by our world-theories - need to have such a past boundary.  

It is needless to say that any remotely plausible world-theory will posit a set $\mathcal{W}$ that is only a minuscule subset of all mathematically definable functions from $\mathcal{M}$ to $E$, in that events occurring at different spatiotemporal locations are expected to hang together in certain intelligible and ideally law-like ways. Indeed, our pre-precise relativistic principle appears to be, to the best of our knowledge, a minimal such ideal restriction that a theory is to obey if it is to have a chance at representing physically possible phenomena. We are now finally ready to give a first articulation of this principle in our framework.

\section*{III. The Principle of Locality: Deterministic}

Our pre-precise principle mandates that an event-occurrence be dependent only on - or be determined entirely by - the events that occur in its immediate causal past. As announced, we will explicate this relation of dependence or determination in terms of physically possible worlds posited by a world-theory. Roughly speaking, we will say that a world-theory is \textit{local}  if, for any point $y$, and past surface $\Sigma$ that is arbitrarily close to $y$, any pair of worlds that agrees on what occurs at $J^-_y\cap \Sigma$ also agrees on what occurs at $y$. Cross-world disagreements on what happens at $y$ ``supervene'' on cross-world disagreements on what happens in the immediate causal past of $y$. Let us state this with more precision.

\bigskip

\textbf{Principle 1.} (\textbf{Locality}) A world-theory $(\mathcal{M},E,\mathcal{W})$ is said to be \textit{local} if it satisfies the following condition. Consider any two surfaces $\Sigma_{t_1}<\Sigma_{t_3}$, and any point $y \in \Sigma_{t_3}$. Then, for any pair of worlds $W,W' \in \mathcal{W}$, there exists a surface $\Sigma_{t_2}$, such that $\Sigma_{t_1}<\Sigma_{t_2}<\Sigma_{t_3}$, and

\begin{equation*}
   \text{if} \quad \left[\forall x \in J^-_y\cap \Sigma_{t_2}: W(x)=W'(x) \right] \quad \text{then} \quad \left[W(y)=W'(y)\right].
\end{equation*}

\bigskip

A theory is thus local iff any two of its possible worlds can disagree on what occurs at $y$ only if they disagree on what occurs in the immediate causal past of $y$. Therefore, even though two worlds may differ regarding what occurs in the \textit{distant }past of $y$, these differences are irrelevant if the worlds turn out to agree on what occurs in the \textit{immediate} past of $y$. The constraint of locality can accordingly be understood as a conjunction of two constraints: (i) an event-occurrence is dependent only on what occurs in its causal past, and (ii) an event-occurrence is in addition entirely determined by events occurring in its \textit{immediate} past. The first component forces event-occurrences to globally respect relativistic causal structure, whereas the second one forces them to influence each other, so to say, only locally in time, thereby making history causally irrelevant. 

It is simple to notice that the principle of locality, as we have now defined it, implies a form of \textit{determinism}. Indeed, any two worlds posited by a local theory can exhibit different events at some point $y \in \Sigma_{t}$ only if they exhibit different events in an arbitrarily close equal-time slice of $J^-_y$. Since the same supervenience condition holds for every element of $\Sigma_{t}$, it follows that two worlds $W,W'$ cannot exhibit different events at $\Sigma_t$, without exhibiting different events at some $\Sigma_{t'}$, with $t'<t$. In other words, any two worlds that agree concerning what happens at some moment in time are bound to also agree on what happens at all later moments. A world-theory can therefore be local only if it is deterministic, that is, only if the events at a surface supervene on the events in an immediately prior surface. On the other hand, determinism does not imply locality, in that an event $y \in \Sigma_t$ may supervene on events at $\Sigma_{t'}$, while not supervening on events in its subset $\Sigma_{t'} \cap J^-_y$. Let us precisify all of this in form of the following definition and proposition.\footnote{The literature on formulations of determinism is large and our simple definition does not do justice to its various intricacies, such as the question of how one is to treat theories that posit worlds enacted by identical quantum particles - see for example Halvorson et al. (2026), and references therein. The definition will however be sufficient for our purposes here, and can be improved in future work if needed.}

\bigskip

\textbf{Definition 2.} (Determinism) A world-theory $(\mathcal{M},E,\mathcal{W})$ is said to be \textit{deterministic} if, for any pair of surfaces $\Sigma_{t_1}<\Sigma_{t_2}$, any point $y\in\Sigma_{t_2}$ and any pair of worlds $W,W' \in \mathcal{W}$: 

\begin{equation*}
   \text{if} \quad \left[\forall x \in \Sigma_{t_1}: W(x)=W'(x) \right] \quad \text{then} \quad \left[W(y)=W'(y)\right].
\end{equation*}

\bigskip

\textbf{Proposition 1.} Locality implies Determinism. Determinism does \textit{not} imply Locality.

\bigskip

All in all, while the here thematized principle can be validated by some deterministic world-theories, it is barred from the very start to possibly be validated by any \textit{in}deterministic world-theory, that is, a world-theory that posits worlds that agree at some time, while disagreeing at later times. Now, quantum phenomena do to the very least appear to be indeterministic, while, to repeat, still retaining a shadow of our relativistic lesson, as manifested, for instance, by the impossibility of the use thereof for superluminal communication. It would thus be valuable to formulate a principle that at least has a \textit{chance} at being validated by quantum phenomena. This is of course what Bell did by proposing his criterion of local causality, albeit concluding that it cannot, up to some qualifications, be upheld by quantum phenomena. We will instead, in the following section, see that a natural generalization of the here analyzed principle leads to a new principle that differs from Bell's, and that does \textit{not} appear to conflict with quantum phenomena after all.

\section*{IV. Probabilistic World-Theories}

Those indeterministic phenomena that are of scientific interest do not manifest their indeterminism in a chaotic or irrational way. If the same conditions gave rise now to this happening, now to that happening, without any intelligible regularity to it, then not much could be said about them, except for something like the following: such and such conditions give rise to \textit{some} happening pertaining to a possibly indefinite collection of possible happenings. Quantum phenomena are, of course, much more orderly than that: given the obtaining of some conditions, (i) there \textit{is} a definite collection of possible happenings that are ideally bound to obtain, and (ii) some of these happenings are more likely to obtain than others, whereby these relative likelihoods are quantifiable in probabilistic terms. For example, given that the spin of an electron is pointing in a certain direction at time $t$, and given that that the electron is to be steered by a magnetic field before impinging on a large screen at time $t'$, the electron (i) is bound to be located at \textit{some} location on the screen at time $t'$, and (ii) it is more likely to be located at some location $\Vec{x}_1$ than at some other location $\Vec{x}_2$, where this relative likelihood is quantifiable in probabilistic terms, as informed by the Born rule. 

World-theories, as introduced in the previous section, can capture only portion (i) - but not portion (ii) - of the above mentioned regularities exhibited by scientifically interesting indeterministic phenomena: they do define the definite collections of possible happenings that are to obtain conditioned on the obtaining of certain conditions, but they do not say anything about which of these are more likely to obtain than others. Let us now state this more precisely. Consider an indeterministic world-theory $(\mathcal{M},E, \mathcal{W})$, that is, a theory which posits worlds that possibly agree on what happens at some surface, without agreeing on what happens in its future. Define now, for any surface $\Sigma$, an equivalence relation `$\sim_{\Sigma}$' on $\mathcal{W}$ as follows:

\begin{equation}
    \left[W \sim_{\Sigma} W'\right] \quad \text{iff} \quad \left[ \forall x \in \Sigma: W(x)=W(x')\right]
\end{equation}

Relation `$\sim_{\Sigma}$' holds between any two worlds that agree on the events that occur on surface $\Sigma$. Accordingly, this relation defines a partition of $\mathcal{W}$ into equivalence classes $\left\{ [W]_{\Sigma}\right\}_{W \in \mathcal{W}}$, where elements within each class agree on what occurs at $\Sigma$, whereas elements pertaining to different classes disagree on what occurs at $\Sigma$. Another way of saying that a theory is deterministic is just to say that the said equivalence classes are all singletons: there are no such $W \neq W'$ that satisfy $W \sim_{\Sigma} W'$, for any $\Sigma$. On the other hand, an indeterministic theory does posit at least one pair of different worlds that, for some $\Sigma$, pertain to the same equivalence class. Moreover, it is the determination of these equivalence classes that captures portion (i) of the regularities pertaining to indeterministic phenomena: each collection of events occurring at some time defines the class of possible worlds within which this collection can figure.

In order to capture portion (ii) of the regularities exhibited by indeterministic phenomena - the relative likelihoods of the obtaining of different worlds - our theories need to be augmented with additional structure, presumably to be formalized in standard probabilistic terms. Now, we are certainly \textit{not} to do this by defining a probability measure over the \textit{whole} set $\mathcal{W}$ of possible worlds, in that there would be no empirical content to such a positing. Indeed, taking two totally different worlds, worlds that for instance do not agree on the events that occur at any location in $\mathcal{M}$ whatsoever: what would it mean to say that one of these worlds is more probable to obtain than the other? What \textit{is} instead empirically meaningful is the likelihood for a world to obtain, \textit{given} that certain conditions obtain: for example, the probability of a coin turning heads in some spatiotemporal region is only empirically meaningful if understood as relativized to, say, the conditions in which the coin has been tossed. Or, inverting the temporal arrow, the probability that it rained yesterday is meaningful only if releativized to, say, the weather conditions that are obtaining right now. 

Instead of assigning a single probability measure to the entire set of worlds $\mathcal{W}$, we will thus assign \textit{multiple} probability measures, each defined on some set of worlds, whose elements are all to share some conditions in common. This ``sharing of conditions'' will here be articulated by an agreement on the events that occur at some conventionally chosen surface $\Sigma_{t_0}$, which can be thought of as hosting the ``initial conditions''. Each probability distribution will thus be defined on equivalence classes $[W]_{\Sigma_{t_0}}$, which we will for simplicity denote as $[W]$, thereby suppressing the index. Note already now that one might also want to speak of distributions defined relative to conditions that do not concern only a single time or a single surface (only the ``initial conditions''), but that refer to multiple surfaces, and thus to multiple different times. It is for example meaningful to speak of the probability that the coin turn heads tomorrow, given that it has been tossed in a certain way now, \textit{and} that it rained yesterday. Such distributions are however, as we will later see, capturable by distributions defined relative only to single surfaces, in terms of ordinary conditional probabilities, thus making an independent introduction thereof redundant.

Taking the above remarks into account, we are ready to introduce \textit{probabilistic world-theories}, that is, world-theories that are augmented by collections of probability measures defined on equivalence classes of worlds. 

\bigskip

\textbf{Definition 3.} (Probabilistic world-theories) A \textit{probabilistic world-theory} is a list $(\mathcal{M}, E, \mathcal{W}, \mathcal{P})$, where $\mathcal{M}$ is a spacetime manifold, $E$ is a set, $\mathcal{W}$ is a set that contains functions $W:\mathcal{M} \rightarrow E$, and $\mathcal{P}\equiv \left\{ \left([W], \mathcal{F}^{[W]}, P^{[W]}\right) \right\}_{W\in\mathcal{W}}$ is a set of probability measures. Each set $[W]$ is the equivalence class of $W \in \mathcal{W}$ defined relative to $\sim_{\Sigma_{t_0}}$, whereas $\mathcal{F}^{[W]}$ is its induced $\sigma$-algebra.  

\bigskip

A probabilistic world-theory is thus a world-theory that is augmented with further structure: besides positing a set of possible worlds, it also posits a set of probability measures, each of which is defined on a subset of worlds that agree on what happens at $\Sigma_{t_0}$, tracked by the equivalence relation $\sim_{\Sigma_{t_0}}$. In other words, such a theory does not merely state that some worlds are physically possible; it also provides, for any two worlds that agree in their initial conditions, the relative likelihood that one of them obtains rather than the other.

It is simple to note that any probabilistic world-theory - in virtue of assigning probability distributions over worlds relative to fixed initial conditions - also induces probability distributions over events possibly occurring in any particular spatiotemporal region, relative to those same initial conditions. Indeed, take a particular probability measure $\left([W], \mathcal{F}^{[W]}, P^{[W]}\right)$ posited by a probabilistic world-theory, defined on a set of worlds that all have the same initial conditions as $W \in \mathcal{W}$. The probability that an event $e \in E$ obtains at some location $x \in M$, relative to the said initial conditions, is given by the marginal distribution

\begin{equation}
    P^{[W]}_x(e)\equiv \sum_{\substack{W' \in [W] \\ W'(x)=e} }P^{[W]}(W'), 
\end{equation}
that is, by the total probability that $e$ obtains at $x$ - relative to the initial conditions shared by the elements of $[W]$ - irrespectively of the events occurring in other spatiotemporal regions. More generally, the joint probability that events $\mathbf{e}=(e_1,...,e_n)$ occur at respective locations $\mathbf{x}=(x_1,...,x_n)$ is given by 

\begin{equation}\label{marginal}
    P^{[W]}_{\mathbf{x}}(\mathbf{e})\equiv \sum_{\substack{W' \in [W] \\ \bigwedge_{i=1}^n W'(x_i)=e_i} }P^{[W]}(W'), 
\end{equation}
where the sum now runs over all those worlds within $[W]$ that feature events $\mathbf{e}$ occurring at $\mathbf{x}$. Stated succinctly, each $P^{[W]}_{\mathbf{x}}(\mathbf{e})$ is the joint probability that events $\mathbf{e}$ occur at $\mathbf{x}$, given that the initial conditions are the ones that are shared among the elements of $[W]$. 

Note that, in the following sections, we will be referring to marginals $P^{[W]}_{\mathbf{x}}$ whose associated set $\mathbf{x}$ is \textit{uncountable} - such as, for instance, a slice $\Sigma \cap J^-_y$ of a past lighcone - thereby pretending that there exists an adequate continuum generalization of Eq. \eqref{marginal}. We take this assumption - which greatly saves on technicalities - to be rather innocuous, since, even if there were, surprisingly, no such adequate continuum generalization, one could nevertheless resort to discretized spacetime models. It is apt to also mention here that it has implicitly been assumed above that the set of worlds $[W]$ is discrete: this assumption is again not conceptually relevant, and can most likely be relaxed with some technical effort, which we will however not pursue here.

\section*{V. The Principle of Locality: Indeterministic}

The previous section introduced probabilistic world-theories, that is, theories that posit possible worlds $\mathcal{W}$ and probability distributions $\mathcal{P}$ over them, relativized to initial conditions that occur at $\Sigma_{t_0}$. As explained, any such theory thereby defines - relative to any initial condition that is shared by the elements of $[W]$ - the joint probability distributions $P^{[W]}_{\mathbf{x}}$ of events possibly occurring in arbitrary spatiotemporal regions $\mathbf{x} \subset \mathcal{M}$. 

We are now ready to proceed with our goal here, which is to explicate the aforementioned relativistic lesson in such a way as for it to be possibly validated by indeterministic phenomena, such as (arguably) the ones studied by quantum theory. We will do so by taking the deterministic principle as a template, and generalizing some of its components so as to apply to probabilistic world-theories. Recall that the deterministic principle states that two worlds cannot differ concerning the event that occurs at some $y$ without differing concerning the events that occur in the immediate causal past of $y$. In other words, according to a local deterministic theory, what happens at $y$ ``supervenes'' on what happens in the immediate causal past of $y$. 

Now, a more general principle - one that has a chance at being validated by some probabilistic world-theories - can be straightforwardly obtained by replacing the kind of agreement that is to enter the said supervenience relation: instead of speaking of the agreement concerning the \textit{event} that occurs at some location, the probabilistic principle is to speak of the agreement concerning the \textit{probability distribution} of events possibly occurring at some location. The new principle is thus to mandate that there cannot be a difference in the distribution of events at $y$ without there being some difference in the distribution of events in the immediate causal past of $y$. In other words, a local probabilistic world-theory is one according to which the \textit{probability} of what happens at $y$ supervenes on the \textit{probability} of what happens in the immediate causal past of $y$. Or, more precisely:

\bigskip

\textbf{Principle 2.} (\textbf{Probabilistic Locality}) A probabilistic world-theory $(\mathcal{M}, E, \mathcal{W}, \mathcal{P})$ satisfies \textit{Probabilistic Locality} - or, more simply, is said to be \textit{local} - if it satisfies the following condition. Consider any two surfaces $\Sigma_{t_1}<\Sigma_{t_3}$, and any point $y \in \Sigma_{t_3}$. Then, for any pair of worlds $W,W' \in \mathcal{W}$, there exists a surface $\Sigma_{t_2}$, such that $\Sigma_{t_1}<\Sigma_{t_2}<\Sigma_{t_3}$, and

\begin{equation*}
   \text{if} \quad \left[P^{[W]}_{\Sigma_{t_2} \cap J^-_y}=P^{[W']}_{\Sigma_{t_2} \cap J^-_y} \right] \quad \text{then} \quad \left[P^{[W]}_{y}=P^{[W']}_{y}\right].
\end{equation*}

\bigskip

According to a local probabilistic world- theory, there cannot be a \textit{probabilistic} difference at $y$, without there being some \textit{probabilistic} difference at an arbitrarily close $\Sigma \cap J^-_y$. The probabilistic and deterministic principles of locality thus manifestly share the same form, and only differ concerning the objects that figure in their respective supervenience relations. Notice furthermore that both principles, in virtue of being formulated in terms of supervenience relations, induce certain \textit{functional} dependencies. Namely, for any location $y$, and slice $\Sigma \cap J^-_y$, a local \textit{deterministic} theory induces a function $F^{(y)}_{\Sigma \cap J^-_y}$ that takes \textit{events} possibly occurring at $\Sigma \cap J^-_y$ as input, and gives the \textit{event} thereby determined to occur at $y$ as output: e.g. the electromagnetic field value at a location is, according to source-free classical electrodynamics, a function of the field values in its immediate causal past. A local \textit{probabilistic} theory induces a functional dependence as well, which however now holds between probability distributions: for any location $y$, and slice $\Sigma \cap J^-_y$, it induces a function $F^{(y)}_{\Sigma \cap J^-_y}$ that takes \textit{probability distributions} of events possibly occurring at $\Sigma \cap J^-_y$ as input, and gives the thereby determined \textit{probability distribution} of events possibly occurring at $y$ as output. The probability of what happens at $y$ is, according to such a theory, a function of the probability of what happens in the immediate causal past of $y$. 

It can be also easily verified that the probabilistic principle \textit{reduces} to the deterministic one in the limiting case of a probabilistic theory that posits only delta-distributions - that is, in the limiting case of a probabilistic theory that is effectively equivalent to a deterministic theory. Indeed, a difference between two delta-distributions of events possibly occurring at some location corresponds to a difference between event-occurrences \textit{simpliciter} occurring at that location. The probabilistic principle therefore appears to be a natural generalization of the deterministic one: (i) it retains the same form, referring to supervenience relations - or functional dependencies - while uniformly substituting distribution-identities for event-identities, and (ii) it reduces to the deterministic principle in case the theory in question is effectively a deterministic one. It is time now to compare our principle to the one proposed by Bell, in order to ultimately establish whether our principle is tenable in light of the empirical violation of Bell's inequalities.

\section*{VI. Probabilistic Locality and Local Causality}

In order to be able to systematically compare our principle of Probabilistic Locality and Bell's principle of local causality, we need to first formulate the latter in the here introduced framework of probabilistic world-theories. Recall that Bell's principle appeals essentially to \textit{conditional probabilities}, in that it tracks how the probability of an event-occurrence changes upon conditioning it on various past conditions. We thus need to first define such conditional probabilities in our framework. The formulation is as expected: letting $P^{[W]}$ be a distribution posited by a probabilistic world-theory, the conditional distribution that events $\mathbf{e} \in E^{|\mathbf{y}|}$ occur at $\mathbf{y}$, conditioned on events $\mathbf{f} \in E^{|\mathbf{x}|}$ occurring at $\mathbf{x}$ - relative to the initial conditions shared by the elements of $[W]$ - is given by
\begin{equation}
    P^{[W]}_{\mathbf{y}|\mathbf{x}}(\mathbf{e}|\mathbf{f})\equiv\frac{P^{[W]}_{\mathbf{y},\mathbf{x}}(\mathbf{e},\mathbf{f})}{P^{[W]}_{\mathbf{x}}(\mathbf{f})}.
\end{equation}

With this definition in place, the below explication of Bell's principle of local causality naturally follows.

\bigskip

\textbf{Principle 3.} (\textbf{Local Causality}) A probabilistic theory $(\mathcal{M}, E, \mathcal{W}, \mathcal{P})$ is said to satisfy \textit{Local Causality} - or, more simply, is said to be \textit{locally causal} - if, for every $W \in \mathcal{W}$, it satisfies the following condition. Consider any $\Sigma_{t_1}<\Sigma_{t_2}$, and any $y \in \Sigma_{t_2}$. Consider also any $z \in \mathcal{M}$, which is spacelike separated from $y$, and which is such that $J^{-}_z \cap (\Sigma_{t_1} \cap J^{-}_y)= \emptyset$. The probability distribution $P^{[W]}$ then satisfies the following relation between its induced conditional probabilities:

\begin{equation*}
   P^{[W]}_{y|\Sigma_{t_1} \cap J^-_y,z}(a|\lambda b)=P^{[W]}_{y|\Sigma_{t_1} \cap J^-_y}(a|\lambda),
\end{equation*}
for arbitrary $a,b \in E$ and $\lambda \in E^{|\Sigma_{t_1} \cap J^-_y|}$.

\bigskip

The latter formulation of Bell's principle has a more cumbersome look than the simpler one presented in section I, owing to our explicit reference to the spatiotemporal locations that the referred events take place at, and to the explicit mention of the worlds $[W]$, that define the initial conditions relative to which the probability distribution is defined. The crux is however the same: according to a locally causal probabilistic world-theory, conditioning on events occurring in the immediate causal past of $y$ fixes the probability distribution of the events that can possibly take place at $y$. Or, in other words, the immediate causal past of $y$ screens off the correlation between events at $y$ and any spacelike separated events that are sufficiently distant from $y$.

Note that Bell's principle exhibits a form that is quite different from the one exhibited by our principle, and thus also from the one exhibited by the deterministic principle. Indeed - as already informally alluded in section I - Bell's principle does \textit{not} appeal to a supervenience relation - or to a functional relation - that is defined on the space of possible phenomena posited by a theory. It instead imposes one constraint separately for each phenomenon modelled by distribution $P^{[W]}$. It accordingly turns out that our Probabilistic Locality and Bell's Local Causality are not only different, but also \textit{logically independent}: the holding of one of them does not, strictly speaking, imply the holding of the other. That this is so can be shown by citing two examples of probabilistic theories, one of which verifies our principle without verifying Bell's, and one that conversely verifies Bell's, but not ours.

The first example is given by a probabilistic world-theory that posits only worlds that \textit{all} agree on what happens at $\Sigma_{t_0}$, and that thus posits only \textit{one} single probability distribution $P^{[W]}$. Our principle of Probabilistic Locality is thereby trivially satisfied, since there is only one distribution posited by the theory: there are \textit{no two} different distributions that could possibly disagree on the probability of some event without disagreeing on the probability of the events in its immediate causal past. Such a theory \textit{can} however obviously contradict Bell's principle, in case this one single distribution $P^{[W]}$ does not satisfy Bell's constraint on its induced conditional marginal distributions. 

The converse example is given by a theory that posits two different equivalent classes of worlds $[W]$ and $[W']$ - and thus two different distributions $P^{[W]}$ and $P^{[W']}$ - which are such that they disagree on the distribution of events at some location $y$, without disagreeing on the distribution of events in the immediate causal past of $y$:

\begin{equation}
    P^{[W]}_{y}\neq P^{[W']}_{y}, \quad  \text{but} \quad P^{[W]}_{\Sigma \cap J^-_y}=P^{[W']}_{\Sigma \cap J^-_y}.
\end{equation}
Such a theory manifestly contradicts our principle; it is however clear that each of the two distributions can \textit{individually }still satisfy Local Causality: whether a single distribution $P^{[W]}$ satisfies Bell's constraint has logically nothing to do with how this distribution relates to other distributions, and thus whether they satisfy our constraint. 

In sum, Bell's principle of Local Causality and our principle of Probabilistic Locality are logically independent, as stated in the following proposition.

\bigskip

\textbf{Proposition 2}. Local Causality does \textit{not} imply Probabilistic Locality. Conversely, Probabilistic Locality does \textit{not} imply Local Causality.

\bigskip

The logical independence of the two principles is not surprising in light of the forms they exhibit: one principle imposes monadic restrictions on each \textit{single} distribution, whereas the other imposes dyadic restrictions on \textit{relations} between distributions. It would however be naive to conclude from the just cited independence, that the empirical violation of Bell's inequalities is only in tension with Local Causality, and not at all with Probabilistic Locality. This would be naive, since there might as well be \textit{further} non-merely-logical principles that might, after all, establish a bridge between ours and Bell's principles. Indeed, even Local Causality does not by itself logically contradict the violation of Bell's inequalities; one needs additional non-merely logical assumptions, whose indubitability is however so stark that it borders with the indubitability of the logical. 

We accordingly need to investigate more deeply the relationship between ours and Bell's principles, by inspecting whether there is some non-merely-logical but hardly-dubitable - or at least plausible - assumption, under the presumption of which our principle implies Bell's principle, and thus ultimately Bell's inequalities. In what follows we will analyze one such plausible assumption, however showing that it can easily be put in doubt, and that the conclusion we ought to draw from the empirical violation of Bell's inequalities is that this additional assumption is in fact false.

\section*{VII. On the Irrelevance of History}

We will present the said additional assumption with the help of a toy example. Suppose that there exists a certain alien type of organisms that can take various \textit{shapes} and \textit{colors}. It is in the nature of such organisms that they tend to change their shapes and colors in regular, probabilistically inspectable manners. In particular, if some such organism, at some moment of time, floats alone in vacuum and looks like a \textit{red pyramid}, then it will spontaneously, after time $\tau$, turn either into a \textit{cube} or into a \textit{sphere}, with $50\%$ chance each. Then, after an additional span of time $\tau$ has passed, this same organism will spontaneously change its color into \textit{blue} or \textit{green}: \textit{if} it was a red \textit{cube} at time $\tau$, then it will turn, deterministically, into a \textit{blue} cube; \textit{if} it was instead a red \textit{sphere}, then it will turn deterministically into a \textit{green} sphere. In sum: red pyramids floating in vacuum turn spontaneously, after time $\tau$, either into red cubes or into red spheres, with $50\%$ chance each. Then, at time $2\tau$, they turn either into blue cubes or into green spheres. 

A probabilistic world-theory that correctly models phenomena involving our alien organisms should, among the possible worlds that it posits, also posit certain worlds whose initial conditions are such that they feature an isolated red pyramid being present in some well localized region. The posited probability distribution over such worlds should then correctly codify the probabilistic changes of shape and color described above. More precisely, letting $[W]$ be the set of worlds that all agree that a red pyramid, immersed in vacuum, is present at some location $x \in \Sigma_{t_0}$, the posited distribution $P^{[W]}$ should satisfy:

\begin{equation}
    P^{[W]}_{x_{\tau},x_{2\tau}}(s,c)=\frac{1}{2}\delta_{c,s}.
\end{equation}
In the latter expression $x_{\tau}$ and $x_{2\tau}$ are the locations at which the shape-change and color-change take place; $s=0,1$ denote the organism being a red cube or a red sphere respectively; and 
$c=0,1$ denote the organism being of blue or green color respectively. Therefore - in accord with what has been described above - if the organism is initially a red pyramid, then it ends up being either a blue cube or a green sphere, with probability $\frac{1}{2}$ each. 

Note now that in the phenomenon under consideration, \textit{if} the organism is a red cube at time $\tau$, then it is bound to turn into a red sphere at time $2\tau$. This feature is captured by our probabilistic theory in form of the following conditional distribution:

\begin{equation}
    P^{[W]}_{x_{2\tau}|x_{\tau}}(c|0)=\delta_{c,0}.
\end{equation}
In other words, if the organism is initially a red pyramid \textit{and} turns into a cube at time $\tau$, then it is bound to turn into a blue cube, in accord with the latter conditional distribution. 

Now, it is widely presumed in physics that history is, in a sense, fundamentally causally irrelevant, or in other words, that fundamental physical dynamics is, in a sense, ``Markovian''. What will happen a few seconds from now - or at least the probability thereof - is determined by what is happening now, regardless of what happened in the past, given that the happenings are described in a sufficiently fine-grained manner. For instance, according to classical mechanics, the configuration of a closed system of particles a few seconds from now is determined by the configuration of particles right now, independently of the past configurations: Laplace's demon, who knows everything about what is occurring right now, does not need to learn anything about the past, in order to perfectly predict the future. Applying this conviction to our toy-example leads one to conclude that red cubes are bound to turn into blue cubes, regardless of their past, regardless of whether they originated from a red pyramid or not. As it happens, those red cubes that we considered above \textit{do} probabilistically originate from red pyramids. This is presumably however wholly contingent: isolated red cubes have an inherent tendency of turning blue, regardless of their history. 

Under the above assumption, the fact that an isolated red cube turns deterministically into a blue cube in \textit{our} phenomenon, implies that isolated red cubes are bound to turn blue in \textit{any} phenomenon whatsoever. A probabilistic theory that complies with this presumption is thus one, according to which, \textit{every} possible world that features, at some moment of time, an isolated red cube, \textit{also} features, after time $\tau$, a blue cube. 
%This applies also to those worlds that feature an isolated red cube as their initial conditions: all such worlds are also to feature a blue cube, after time $\tau$.
Consequently, if the probability of there being an isolated red cube at some moment of time is non-zero, then, \textit{conditioned} on this possibility, the probability of there being a blue cube after time $\tau$ should be 1, and this is to hold regardless of the history that might precede the red cube. Generalizing beyond our toy-example, for any way the world could be at some moment - for any conglomerate of events occurring at a surface - the probability distribution over the ways the world can subsequently unfold is independent of what actually unfolded \textit{before} the said moment. 

We will now turn the above assumption - the assumption of the ``irrelevance of history'' - into a precise constraint on probabilistic world-theories. Let $\mathbf{e}^* \subset E^{|\Sigma|}$ be a collection of events that, according to a probabilistic world-theory, can possibly occur at a surface $\Sigma$. The conglomeration of events $\mathbf{e}^*$ is thus a way the world could be at at an arbitrary moment in time, according to the theory in question. Let us for simplicity introduce notation $W(\Sigma)$ to denote the list of events that function $W$ assigns to the elements of $\Sigma$:
\begin{equation}
    W(\Sigma)\equiv(W(x))_{x \in \Sigma}
\end{equation}
The equivalence class of worlds that all agree that events $\mathbf{e}^*$ occur at surface $\Sigma_{t_0}$ is then given by
\begin{equation}
    [W]_{\mathbf{e}^*}\equiv \left\{W \in \mathcal{W}|W(\Sigma_{t_0})=\mathbf{e}^* \right\}.
\end{equation}

Consider also \textit{another }collection of worlds $[V]$, which are such that, relative to their initial conditions, the probability of the said events $\mathbf{e}^*$ to occur at a later surface $\Sigma >\Sigma_{t_0}$ is non-zero: $P^{[V]}_{\Sigma}(\mathbf{e}^*)\neq 0$. In other words, the initial conditions shared by the worlds in $[V]$ are such that the said events $\mathbf{e}^*$ \textit{can} arise at some later moment. Now, denote with $[V]^{(\Sigma)}_{\mathbf{e}^*}\subset [V]$ the subset of worlds that contains all and only those worlds in $[V]$ that all agree that $\mathbf{e}^*$ does occur at $\Sigma$, that is:

\begin{equation}
    [V]^{(\Sigma)}_{\mathbf{e}^*}\equiv \left\{V' \in [V]|V'(\Sigma)=\mathbf{e}^* \right\}.
\end{equation}

Both classes $[W]_{\mathbf{e}^*}$ and $[V]^{(\Sigma)}_{\mathbf{e}^*}$ thus feature events $\mathbf{e}^*$ at some surface: the substantial difference is only that the latter class, but not the former one, is furthermore constrained to also feature events $V(\Sigma_{t_0})$ at some prior surface. That the theory in question regards history as irrelevant then means, roughly, that the theory treats the futures of events $\mathbf{e}^*$ in the same way relative to the two classes: i.e., it treats the futures of these events as independent of whether the initial conditions are given by events $\mathbf{e}^*$ themselves, or if they are given by some other events $V(\Sigma_{t_0})$ that precede events $\mathbf{e}^*$. In other words, the probability of some event occurring time $\tau$ after surface $\Sigma$ - conditioned on events $\mathbf{e}^*$ having occurred at $\Sigma$ \textit{and} given that the initial conditions at $\Sigma_{t_0}$ are the ones shared by $[V]$ - equals the probability of such an event occurring time $\tau$ after $\Sigma_{t_0}$ - given that the initial conditions at $\Sigma_{t_0}$ are given by events $\mathbf{e}^*$ themselves. Applying this to our toy-example, the probability of there being a blue cube after there being an isolated red cube is independent of whether the red cube originated from a red pyramid or not. Or, more precisely:

\bigskip

\textbf{Principle 4.} (\textbf{Irrelevance of History}) A probabilistic world-theory $(\mathcal{M}, E, \mathcal{W}, \mathcal{P})$ is said to satisfy \textit{Irrelevance of History} - or, more simply, is said to regard history as irrelevant - if it satisfies the following constraint. Let $\mathbf{e}^* \subset E^{|\Sigma|}$ be a collection of events that can possibly occur at a surface, and let $[W]_{\mathbf{e}^*}\equiv \left\{W \in \mathcal{W}|W(\Sigma_{t_0})=\mathbf{e}^* \right\}$. Let $[V]\subset \mathcal{W}$ be such that $P^{[V]}_{\Sigma}(\mathbf{e}^*)\neq 0$ holds, for some surface $\Sigma>\Sigma_{t_0}$, and let $[V]^{(\Sigma)}_{\mathbf{e}^*}\equiv \left\{V' \in [V]|V'(\Sigma)=\mathbf{e}^* \right\}$. Then there exists a function $f: [W]_{\mathbf{e}^*} \rightarrow [V]^{(\Sigma)}_{\mathbf{e}^*}$, such that, for all $W \in [W]_{\mathbf{e}^*}$:\\
\smallskip
(i) $[f(W)](x)=W(T_{\Sigma}^{-1}(x))$, for all $x \in \left(J^{+}_{\Sigma} \cup \Sigma \right)$ and\\
(ii) $P^{[V]}(f(W))= P^{[V]}_{\Sigma} (\mathbf{e}^*)P^{[W]_{\mathbf{e}^*}}(W)$,\\
\smallskip
where $T_{\Sigma}: \mathcal{M} \rightarrow \mathcal{M}$ is a function that translates points in $\Sigma_{t_0}$ to points in $\Sigma$.

\bigskip

Let us unpack this. Function $f$ occurring above associates to each world $W \in [W]_{\mathbf{e}^*}$ - with initial conditions $\mathbf{e}^*$ - a world $f(W) \in [V]^{(\Sigma)}_{\mathbf{e}^*}$ - whose initial conditions are \textit{not} given by $\mathbf{e}^*$, but that features these same events at some later surface $\Sigma$. According to (i), this association is such that each $f(W)$ is a spatiotemporally translated version of $W$: if, according to $W$, a red cube is present at some $x$ that is either at $\Sigma_{t_0}$ or in its future, then, according to $f(W)$, a red cube is present at $T_{\Sigma}(x)$, which is located either at $\Sigma$ or in its future. Furthermore, according to (ii), the probability distribution over worlds in $[W]_{\mathbf{e}^*}$ is equal to the probability distribution over $f$-related worlds in $[V]^{(\Sigma)}_{\mathbf{e}^*}$, up to the normalization constant $P^{[V]}_{\Sigma} (\mathbf{e}^*)$. A theory that regards history to be irrelevant thus, in accord with what we have said so far, does \textit{not }discriminate between (a) the distribution over the possible futures that might arise out of events $\mathbf{e}^*$ occurring at $\Sigma_{t_0}$, and (b) the distribution over the possible futures that might arise out of these same events $\mathbf{e}^*$ occurring at some later surface $\Sigma>\Sigma_{t_0}$. That is, \textit{given} that events $\mathbf{e}^*$ occur, the distribution over the possible futures is fixed, and is independent of what might have occurred \textit{before} $\mathbf{e}^*$. 

Recall now that the reason we have digressed into talking about the irrelevance of history was to investigate the relationship between our principle of Probabilistic Locality and Bell's principle of Local Causality. Indeed, it turns out, as we will now briefly explain, that, \textit{if} history is irrelevant, then our principle \textit{implies} Bell's principle. Consider namely a probabilistic theory that does \textit{not} satisfy Bell's principle. Such a theory posits at least one distribution $P^{[W]}$, which is such that 

\begin{equation}\label{nonloc}
   P^{[W]}_{y|\Sigma \cap J^-_y,z}(a|\lambda b)\neq P^{[W]}_{y|\Sigma \cap J^-_y,z}(a|\lambda b'),
\end{equation}
where $y,z$ and $\Sigma$ satisfy the properties that enter into the definition of Local Causality. The distribution at $y$ is thus not fixed upon conditioning on events $\lambda$ in its immediate causal past; it is instead also sensitive to conditioning on whether $b$ or $b'$ occurs at a spacelike separated location $z$. Let us now, without loss of generality, assume that the location $z$, at which $b$ or $b'$ occur, is an element of the same surface at which $\lambda$ occur, i.e. $z \in \Sigma$.\footnote{The reason for which no generality is lost here is that $b$ (or $b'$) is assumed - per Local Causality - to occur at spacelike separation from all the events in $\lambda$: there is thus an inertial reference frame - associated to some foliation - for which $b$ (or $b'$) and $\lambda$ lie on the same equal-time surface.}. Let us denote with $\mathbf{e} \in E^{|\Sigma|}$ a particular conglomerate of events that can occur at surface $\Sigma$, which comprises events $\lambda$ occurring at $\Sigma \cap J^-_y$, event $b$ occurring at $z$, and some other arbitrary events occurring at the remaining locations in $\Sigma$. Similarly, let us denote with $\mathbf{e}' \in E^{|\Sigma|}$ a conglomerate of events that can occur at $\Sigma$, which also comprise events $\lambda$ occurring at $\Sigma \cap J^-_y$, but that comprise event $b'$ occurring at $z$, and some further arbitrary events occurring at the remaining locations in $\Sigma$. It then holds, in accord with Eq. \eqref{nonloc}, that 
\begin{equation}\label{nonloc2}
   P^{[W]}_{y|\Sigma}(a|\mathbf{e})\neq P^{[W]}_{y|\Sigma}(a|\mathbf{e}').
\end{equation}

Now, consider the collection of worlds whose initial conditions are given by $\mathbf{e}$, that is $[W]_{\mathbf{e}}\equiv \left\{W \in \mathcal{W}|W(\Sigma_{t_0})=\mathbf{e} \right\}$. The principle of the Irrelevance of History then implies that there ought to be a strong tie between the distribution of worlds in $[W]_{\mathbf{e}}$, and the distribution of those worlds in $[W]$ that feature events $\mathbf{e}$ at $\Sigma$. In particular, it ought to hold that
\begin{equation}\label{ass}
    P^{[W]}_{y|\Sigma}(a|\mathbf{e})=P^{[W]_{\mathbf{e}}}_{T^{-1}_{\Sigma}(y)}(a),
\end{equation}
i.e., that the probability that $a$ occurs after $\mathbf{e}$ occurred is independent of what occurred before $\mathbf{e}$. The analogous conclusion is also reached concerning the other conglomerate of events $\mathbf{e}'$:
\begin{equation}\label{ass2}
    P^{[W]}_{y|\Sigma}(a|\mathbf{e}')=P^{[W]_{\mathbf{e}'}}_{T^{-1}_{\Sigma}(y)}(a),
\end{equation}
where $[W]_{\mathbf{e}'}\equiv \left\{W \in \mathcal{W}|W(\Sigma_{t_0})=\mathbf{e}' \right\}$. Conjoined with Eq. \eqref{nonloc2}, it then trivially follows that 
\begin{equation}\label{nonloc3}
    P^{[W]_{\mathbf{e}}}_{T^{-1}_{\Sigma}(y)}(a) \neq P^{[W]_{\mathbf{e}'}}_{T^{-1}_{\Sigma}(y)}(a).
\end{equation}

We are faced here with two different initial conditions $\mathbf{e}$ and $\mathbf{e}'$, which determine two different probabilities that event $a$ occurs in the future of $\Sigma_{t_0}$. Importantly, the two conditions do not at all differ concerning the events that occur at the intersection between $\Sigma_{t_0}$ and the causal past of event $a$, i.e. they do not differ concerning what happens at $\Sigma_{t_0} \cap J^{-}_{T^{-1}_{\Sigma}(y)}$. 

Now, recall that our principle of Probabilistic Locality dictated that probabilistic differences at any $y$ supervene on probabilistic differences in the immediate causal past of $y$. Supposing that $y$ is located in the future of $\Sigma_{t_0}$, an iteration of the same condition for the points in $J^-_y$ implies that there cannot be a probabilistic difference at $y$ without there being some probabilistic difference at $\Sigma_{t_0} \cap J^{-}_y$. Since each probability distribution posited by a probabilistic world-theory is defined on an equivalence class of worlds that agree on what occurs at surface $\Sigma_{t_0}$, the distribution of the events at this surface is necessarily a delta function. In other words, it holds, by definition, that $P^{[W]}_{\Sigma_{t_0}}(W(\Sigma_{t_0}))=1$. Therefore, there cannot be a probabilistic difference at $y$ without there being a difference \textit{simpliciter} at $\Sigma_{t_0} \cap J^-_y$. Fixing the initial conditions fixes the probability distribution of the events that possibly occur in their future, and it does so while respecting relativistic lightcone structure. 

The reason we have pointed out this particular consequence of Probabilistic Locality is that it is flatly \textit{contradicted} by our example above: according to Eq. \eqref{nonloc3}, there \textit{is} a difference in the probability distribution of events possibly occurring at $T^{-1}_{\Sigma}(y)$, \textit{without} there being a difference in the events occurring at $\Sigma_{t_0}\cap J^{-}_{T^{-1}_{\Sigma}(y)}$. Therefore, \textit{if} Irrelevance of History holds, then any probabilistic world-theory that violates Bell's principle of Local Causality \textit{also} violates our principle of Probabilistic Locality:

\bigskip

\textbf{Proposition 3}. Probabilistic Locality and Irrelevance of History jointly imply Local Causality.

\bigskip

The tension between the empirical violation of Bell's inequalities and Local Causality consequently translates into an equal tension with our Probabilistic Locality, \textit{unless} the assumption concerning the Irrelevance of History is false. Given that we have good empirical reasons to hold on to our principle of Probabilistic Locality, and that we seem not to have good reasons for doubting any of the other assumptions that implicitly or explicitly enter Bell's derivation of his inequalities - such as the assumption of  ``statistical independence'' - we seem to be given no other choice than to accept that the aforementioned assumption is false, and that history is relevant after all. Let us now inspect the coherence of this proposal, and draw some of its consequences. 

\section*{VIII. On the Relevance of History}

As noted in the previous section, the empirical violation of Bell's inequalities, conjoined with the empirical support of relativity theory, gives us reason to hold on to our principle of Probabilistic Locality, while rejecting the principle that mandates that history be irrelevant. Let us now better understand what it is for history to be \textit{relevant}, by returning to our toy-example, the one that spoke of our alien organisms taking on various shapes and colors. Recall that any red cube, as long as it originated from a red pyramid, is bound to turn deterministically into a blue cube. If we now accept that history may be relevant, then we ought \textit{not} to \textit{a priori} regard the origin of the red cube as immaterial to its future and thereby conclude that any red cube whatsoever is bound to turn blue. It might instead be the case that the possible future developments of a red cube depend non-trivially on its different possible origins. Our alien organisms might, for example, exhibit the following behavior. \textit{If} a cube originated from a red \textit{pyramid}, then, as already said, it is bound to turn blue. However, \textit{if} the cube originated, say, from a red \textit{sphere}, then it is bound to keep being red. Moreover, washing out their possible origins, red cubes \textit{on average}, say, either turn blue or keep being red, with $50\%$ chance each. 

If our alien organisms behave as sketched above, then, upon finding an isolated red cube, we ought not to conclude that it will certainly turn blue later. We ought to instead place equal bets on it turning blue or on it keep being red, since this is how red cubes behave on average, irrespective of their origins. However, if we learn that the particular red cube that we are pondering upon in fact originated from a red pyramid, then we ought to update our beliefs concerning the future, and place all our bets on the cube turning blue. More generally, every single red cube that we might ever encounter does \textit{in fact} lie at the tip of one particular indefinitely long history, one among the many possible histories that may have brought forth red cubes on their tips.\footnote{Note that if there were instead only \textit{one} possible history that could have preceded our red cube, then it would \textit{trivially} follow that history is irrelevant: there would be no room for learning something about history, since the red cube's entire history could, at least in principle, be inferred from its mere presence. This applies also to time-symmetric deterministic theories, such as Newtonian mechanics, for which the present not only determines the future, but also the past: such theories leave no room for history to be relevant, as they deem it to be entirely determined by the present.} While the future development of a red cube under examination may non-trivially depend on the particular history that preceded it, this does \textit{not} imply that being ignorant about the latter precludes us from knowing at least something about the future of the red cube in question. What the relevance of history instead \textit{does} imply is that there is in principle always room for \textit{improvement} of our predictions concerning the future: learning more and more about actual history may keep refining our expectations of the future, while certainly not contradicting these expectations.\footnote{Learning that the red cube in question did in fact originate from a red pyramid does not lead us to contradict our prior expectations, but only to refine them: at first we hypothesize that the red cube is to turn blue with $50 \%$ chance, but then, upon learning about its history, we get convinced that it will turn blue with $100 \%$ chance. This is coherent. It would instead be \textit{in}coherent if, say, red cubes on average turned blue with $0 \%$ chance, but, in case they originated from a red pyramid, turned blue with some non-zero chance.} 

Let us now make a further remark on the relevance of history. Take again that we are confronted with a particular red cube, whose actual origins are unknown to us. Not knowing its origins also implies that we do not know \textit{how likely} it is for the red cube to be there, where it actually is, relative to its past origins. On the other hand, if we were to learn about its origins, then we would also learn something about the likelihood of the red cube being present there where it is, together with the likelihoods of alternative ways in which history might have unfolded, but happened not to. History being relevant for the determination of the future is thus compatible with currently unrealized possibilities and their pertaining likelihoods \textit{also} being relevant for the determination of the future. While what is currently happening only partially determines the distribution over possible futures, it is compatible with the relevance of history that this determination be refinable by taking note, not only of the history of what actually happened, but also of what merely \textit{could} be \textit{currently} happening, together with the likelihoods of these merely possible current happenings, relative to their origins. 

The just noted possibility that the distribution of current happenings might be relevant in the determination of the future is exactly what is fulfilled by our principle of Probabilistic Locality. Indeed, our principle stated that the probability distribution of what might happen at some $y$ is entirely determined by the probability distribution of what might have happened in the immediate causal past of $y$. It is not what \textit{actually} unfolded in the past vicinity of $y$ that entirely determines the distribution of what goes on at $y$, but what \textit{could} have unfolded, together with the distribution of these possible unfoldings. It is consequently also clear that our principle of Probabilistic Locality bears a shadow of the \textit{ir}relevance of history, since what matters in the determination of the distribution of $y$ is only the distribution of what happens in the immediate causal past of $y$, \textit{irrespective} of the distribution of possible happenings in the more distant past. Therefore, while what is happening right now only partially determines the distribution of possible future happenings - and while this determination can be refined by citing actual past happenings - it is still the case that the \textit{distribution} of what \textit{might} be happening right now \textit{does} entirely determine the distribution of possible future happenings. History is thus relevant in one sense and irrelevant in another: the history of what actually happened \textit{is} relevant; the history of what could have happened, together with their distributions, is \textit{not}.

It is thus coherent to maintain history to be relevant - in the sense of denying the assumption of the Irrelevance of History - while maintaining that relations of determination are, in a certain sense, local, as manifested in form of our principle of Probabilistic Locality. In other words, there are probabilistic world-theories that validate Probabilistic Locality, while simultaneously invalidating the Irrelevance of History. The empirical violation of Bell's inequalities and the empirical support for relativity theory thus arguably indicate that, if we are to make sense of quantum phenomena, then we are to accept the validity of some such probabilistic world-theory. Let us now, for clarity, briefly make explicit the way in which such a local theory can invalidate Bell's principle, and thus be compatible with the violation of his inequalities. Referring to the discussion in the previous section, it is clear that such a theory - if it is to comply with Probabilistic Locality - ought to deny either Eq. \eqref{ass} or Eq. \eqref{ass2}, and thus assert, for instance: 
\begin{equation}\label{ass3}
    P^{[W]_{\mathbf{e}}}_{T^{-1}_{\Sigma}(y)}(a)\neq P^{[W]}_{y|\Sigma}(a|\mathbf{e}).
\end{equation}
This equation states the inequivalence of two probabilities. The probability appearing on the LHS is the probability that event $a$ occurs some time after events $\mathbf{e}$ have occurred, \textit{independently} of what occurred before $\mathbf{e}$, averaging out what might have occurred beforehand. The RHS on the contrary speaks of the probability that event $a$ occurs some time after events $\mathbf{e}$ have occurred, \textit{given} that events $\mathbf{e}$ were preceded by events $W(\Sigma_{t_0})$, i.e. by the initial conditions shared by the worlds in $[W]$. Eq. \eqref{ass3} is thus a paradigmatic exemplification of the relevance of history: the probability that $a$ occurs depends on the history of events $\mathbf{e}$, on whether they are preceded by events $W(\Sigma_{t_0})$ or not. And it is precisely the biting of this bullet that blocks the transformation of a violation of Local Causality into a violation of Probabilistic Locality. It is thus coherent to take Bell's inequalities to contradict Local Causality, without taking them to contradict our principle of Probabilistic Locality.

The above reflection clarifies the relationship between the two principles of locality; nevertheless, one might still have some doubts about how exactly these abstract considerations apply to \textit{concrete} experiments in which Bell inequalities can actually be violated. What is the exact way in which EPR-like experiments can be taken to invalidate Bell's principle, and yet validate our principle? An adequate answer to this would require the construction of concrete theories that contain plausible models of EPR-like experiments, and which violate Bell's principle, but satisfy ours. While this constructive project lies much beyond our scope here, we will nevertheless now briefly indicate the core rationale lying behind any such theory's account of Bell.

Recall that Bell-inequality-violating EPR-like experiments feature two spacelike separated measurements executed on a pair of particles initially prepared in some entangled state $\psi$. These measurements enact Bell-inequality-violating distributions $P_{\psi}^{(r,s)}(a,b)$, where $(a,b)$ are the measurements' outcomes and $(r,s)$ their settings. Our goal now is to cursorily inspect the mechanism by which a probabilistic world theory $\mathcal{T}$ can account for such experiments, all the while complying with our principle of locality. First, $\mathcal{T}$ ought to, of course, provide empirically adequate models of the said experiments. In order to do so, it ought - for each pair of measurement settings $(r,s)$ - posit an equivalence class of worlds $[W]^{r,s}_\psi$, whose initial conditions all feature measurement devices being configured according to settings $(r,s)$, and a pair of particles being in state $\psi$. The corresponding distributions $P^{[W]^{r,s}_\psi}$ posited by $\mathcal{T}$ ought to then satisfy, for each $(r,s)$:
\begin{equation}
    P_{x_a,x_b}^{[W]^{r,s}_\psi}=P_{\psi}^{(r,s)},
\end{equation}
where $x_a,x_b \in J^+_{\Sigma_0}$ denote the spatiotemporal locations at which events $a,b$ occur. 

Focus now on an intermediate surface $\Sigma$, that lies in the future of the initial conditions, but in the past of, say, measurement outcome $a$. Take two conglomerates of events $\mathbf{e}, \mathbf{e'} \in E^{|\Sigma|}$ that $\mathcal{T}$ regards as possibly occurring at surface $\Sigma$. Assume in particular that $\mathbf{e}$ and $\mathbf{e'}$ \textit{fully agree} on what happens at $\Sigma \cap J^-_{x_a}$, but disagree elsewhere. Finally, suppose that, for some pair of settings $(r,s)$, the said distribution $P^{[W]^{r,s}_\psi}$ satisfies the following relation
\begin{equation}\label{nonlocs}
    P^{[W]^{r,s}_\psi}_{x_a|\Sigma}(a|\mathbf{e}) \neq P^{[W]^{r,s}_\psi}_{x_a|\Sigma}(a|\mathbf{e'}).
\end{equation}

Theory $\mathcal{T}$ thus clearly violates Bell's principle of Local Causality, since the probability of outcome $a$ is not fixed by conditioning on what occurs in a slice $\Sigma \cap J^-_{x_a}$ of its causal past. Nevertheless, $\mathcal{T}$ need \textit{not} violate our principle of Probabilistic Locality, since it can still, via a violation of the Irrelevance of History, maintain that the following holds:
\begin{equation}\label{locexx}
    P^{[W]_{\mathbf{e}}}_{T^{-1}_{\Sigma}(x_a)}(a) = P^{[W]_{\mathbf{e'}}}_{T^{-1}_{\Sigma}(x_a)}(a),
\end{equation}
where $[W]_{\mathbf{e}}$ and $[W]_{\mathbf{e'}}$ are classes of worlds, whose initial conditions are respectively given by $\mathbf{e}$ and $\mathbf{e'}$. Therefore, the non-locally-appearing correlations in Eq. \eqref{nonlocs} are, according to $\mathcal{T}$, \textit{washed out} by averaging over the possible histories that could have lead to $\mathbf{e}$ or to $\mathbf{e'}$, thereby validating the form of locality exhibited in Eq. \eqref{locexx}.\footnote{This is of course only coherent if $\mathcal{T}$ posits that multiple possible histories can lead to the very same $\mathbf{e}$, or to the very same $\mathbf{e'}$ - a feature that is perfectly expected by an indeterministic theory. In fact, this feature is \textit{necessarily} satisfied by any \textit{time-symmetric} indeterministic theory: the present can give rise to many possible futures, and so symmetrically, the present could have also come about from many possible histories.} In other words, the theory tells us that the apparently nonlocal element that we encounter in EPR-like experiments is a mere consequence of our contingent focus on a particular subset of histories that could have lead to $\mathbf{e}$ or to $\mathbf{e'}$, namely, on the subset of histories that feature the corresponding EPR-like initial conditions at some surface prior to $\Sigma$. This is, in the roughest outline, the rationale behind any local probabilistic world-theory that aspires to digest Bell-inequality-violating phenomena by accepting the relevance of history. Whether some such plausible theory can actually be constructed is another story.

\section*{IX. A Consequence of Time Symmetry}

The previous section has shown that accepting history to be relevant allows a probabilistic world-theory to coherently admit the violation of Bell's inequalities and invalidate the old principle of Local Causality, all the while obeying the lessons of relativity, codified in form of our new principle of Probabilistic Locality. In this section we will briefly indicate a further consequence of our principle, a consequence that follows by conjoining it with the apparent \textit{temporal symmetry} of fundamental physical phenomena. That fundamental physical phenomena appear to be temporally symmetric means that the physical possibility of any such phenomenon appears to imply the physical possibility of its temporal inverse. In virtue of its generality - it not concerning the nature of the events that figure in the said phenomena - the principle of temporal symmetry can be straightforwardly incorporated within our general framework, as follows.\footnote{See Price (1996) for a classic reflection on issues related to time-symmetry in physics. For a more recent overview, see Callender (2026) and references therein.} 

In order to state what it is for a world-theory to be time-symmetric, we need to first define the temporal inverse of a world posited by any such theory. This in turn requires reference to the temporal inversion of the world at any moment in time: for example, the temporal inverse of a classical particle being positioned at $\Vec{x}$ and having velocity $\Vec{v}$ should be a particle of the same type also being positioned at $\Vec{x}$ but having velocity $-\Vec{v}$. We will not assume any particular properties of the function that codifies such a temporal inversion of a world at a moment of time, but we will merely represent it with function $T: E^{|\Sigma|} \rightarrow E^{|\Sigma|}$. Therefore, for any way $\mathbf{e} \in E^{|\Sigma|}$ that the world could be at any moment of time - at any surface $\Sigma$ - the temporal inversion thereof is given by $T(\mathbf{e})$. With this primitive in hand, we can easily define the temporal inversion of any world as a whole. Indeed, $\tau(W)$ is the temporal inverse of any world $W \in \mathcal{W}$ posited by a (probabilistic) world-theory, if it satisfies:
\begin{equation}
    [\tau(W)](\Sigma_t)=T[W(\Sigma_{-t})],
\end{equation}
for any surface $\Sigma_t$. In other words, $\tau(W)$ is the temporal mirror image of $W$, up to the above introduced function $T$. 

With the time-inversion function $\tau$ in our hands, we are now ready to define what it is for a world-theory to be \textit{time-symmetric}, starting with the \textit{non-}probabilistic case. A world-theory is namely time-symmetric if it satisfies
\begin{equation}
    W \in \mathcal{W} \quad \text{iff} \quad \tau(W) \in \mathcal{W},
\end{equation}
that is, if the theory regards a world to be physically possible if and only if it also regards its temporal inverse to be physically possible. 

%Recall now that a \textit{deterministic} world-theory is one that does not allow for any pair of worlds to agree on what happens at a Cauchy surface without agreeing in their future events. It is simple to see that a \textit{time-symmetric deterministic} theory is accordingly one that does not allow for any pair of worlds to agree on what happens at a Cauchy surface without agreeing in the past of that surface: as much as the same conditions cannot give rise to different futures, so also the same conditions cannot arise from different pasts. Conversely, a time-symmetric \textit{in}deterministic world-theory is one that not only allows for the same conditions to give rise to different futures, but that also allows the same conditions to arise from different pasts. The future is determinate if and only if the past is determinate.

\textit{Probabilistic }world-theories, in addition to the above constraint on the posited sets of worlds $\mathcal{W}$, ought to also incorporate a further related constraint on their posited sets of probability distributions $\mathcal{P}$. That is, they should also validate the following tie between the distribution $P^{[W]}$ over worlds in $[W]$ and the distribution $P^{[\tau(W)]}$ over the corresponding temporally inverted worlds in $[\tau(W)]$:
\begin{equation}
    P^{[\tau(W)]}(W')=P^{[W]}(\tau^{-1}(W')).
\end{equation}
A theory that does \textit{not} validate the latter would indeed break temporal symmetry at the statistical level: while the mere occurrence of $\tau(W)$ would not betray the temporal arrow, the empirically inspectable \textit{probability} of its occurrence, relative to its initial conditions, \textit{would} betray it. 

Now, recall that a probabilistic world-theory is \textit{local} if any probabilistic difference at some location $y$ supervenes on probabilistic differences in the immediate causal past of $y$. It is simple to verify that, if such a theory is also \textit{time-symmetric}, then it satisfies a further principle, the time-inverted version of Probabilistic Locality. More precisely, a probabilistic world-theory $(\mathcal{M}, E, \mathcal{W}, \mathcal{P})$ that is both local and time-symmetric also satisfies the following condition. Consider any two surfaces $\Sigma_{t_1}<\Sigma_{t_3}$, and any point $y \in \Sigma_{t_1}$. Then, for any pair of worlds $W,W' \in \mathcal{W}$, there exists $\Sigma_{t_2}$, such that $\Sigma_{t_1}<\Sigma_{t_2}<\Sigma_{t_3}$, and

\begin{equation*}
   \text{if} \quad \left[P^{[W]}_{\Sigma_{t_2} \cap J^+_y}=P^{[W']}_{\Sigma_{t_2} \cap J^+_y} \right] \quad \text{then} \quad \left[P^{[W]}_{y}=P^{[W']}_{y}\right].
\end{equation*}

In other words, any probabilistic difference at some $y$ supervenes on probabilistic differences in the immediate causal \textit{future} of $y$. Note that the obtaining of this supervenience relation by no means implies any form of ``retrocausality'', according to which current events can supposedly causally influence past events. My current presence in my room implies or determines my presence in this room a few milliseconds ago, but this by no means amounts to saying that my current state ``causally influences'' my past states. In order to avoid confusion, it would thus be more advisable, instead of speaking of the determination of the present on behalf of the future, to speak contrapositively of the necessity that any event leave an imprint in its future: any probabilistic difference here-and-now is to make a probabilistic difference in the immediate causal future. According to a local time-symmetric probabilistic world-theory, there can be no traceless probabilistic difference, as much as - according to a time-symmetric \textit{deterministic} world-theory - there can be no traceless difference \textit{simpliciter}.\footnote{According to a time-symmetric deterministic world-theory, not only does the present determine the future, but it also determines the past: for any way the world could be at some $\Sigma$, there is only one way in which the world could have been before $\Sigma$, and only one way in which it can unfold thereafter.} 

As we saw beforehand, an empirically adequate probabilistic world-theory can respect locality only by deeming history to be relevant. If such a theory is also time-symmetric, then, as it can easily be inspected, the relevance of history also implies the \textit{relevance of the future}: as much as the probability that it rains tomorrow may depend both on what is occurring now and on what occurred yesterday, so also the probability that it rained yesterday may depend not only on what is occurring now, but also on what will occur tomorrow. More precisely, such a theory may regard the two following probabilities to be inequivalent: (a) the probability that $a$ occurs at some location $y$ in the past of $\Sigma_{t_0}$, relative to some initial conditions defined on $\Sigma_{t_0}$, and (b) the probability that $a$ occurs at the same location $y$, relative to the same initial conditions, but conditioned on certain events $\mathbf{e}$ occurring at a later surface $\Sigma>\Sigma_{t_0}$. In other words, the following relation may hold:
\begin{equation}\label{ass4}
    P^{[W]}_{y|\Sigma}(a|\mathbf{e}) \neq P^{[W]_{\mathbf{e}}}_{T^{-1}_{\Sigma}(y)}(a)
\end{equation}
The sole difference between this relation - which codifies the relevance of the future - and the previous Eq. \eqref{ass3} - which codified the relevance of history - is the temporal order that holds between the referred events and surfaces: event $y$ is now located in the past of $\Sigma_{t_0}$, which is in turn located in the past of $\Sigma$. 

Summing up, a time-symmetric local probabilistic world-theory - besides regarding probabilistic differences as supervenient on probabilistic differences in the causal past, and besides deeming history to be relevant - \textit{also} regards probabilistic differences as supervenient on probabilistic differences in the causal future, and deems the future to be relevant. In so far as fundamental physical phenomena allow for the violation of Bell's inequalities, respect relativistic constraints, \textit{and} are time-symmetric, only a theory of the just cited sort has a chance at being adequate.

\section*{X. What have we done? What is still to be done?}

Let us summarize the proposal that has been presented in the preceding sections, before contextualizing it within the existing literature. We saw at the outset that Bell's principle of local causality does not appear to be an adequate probabilistic generalization of its deterministic variant. It is \textit{another} principle that we should instead adhere to, one that is explicitly formulated in terms of supervenience relations - or functional dependencies - and that reduces to the deterministic case in its relevant limit. This new principle of Probabilistic Locality however turns out to imply Bell's old principle, \textit{if} it is additionally mandated that history be irrelevant. Our proposal is thus to accept the possibility that history is \textit{relevant}, thereby enabling a probabilistic world-theory to allow for the violation of Bell's inequalities, while nevertheless validating our new principle of locality. Such a theory may, as finally noted, also be time-symmetric, thus inducing the temporally inverted renditions of Probabilistic Locality and of the relevance of history. 

The conclusion that we have drawn from the empirical violation of Bell's inequalities differs from many of the conclusions that have been drawn in the literature so far. Our proposal implies that it was premature to conclude, as some have, that locality cannot be maintained in light of the empirical violation of Bell's inequalities.\footnote{For a detailed and forceful push for taking Bell's work to imply non-locality, see Maudlin (2011).} It \textit{can} be maintained, given that its probabilistic rendition is adequately understood. We are also parting ways from those that reject the assumption of statistical independence, whose only reason for doing so seems to be the finding of no other way out in dealing with Bell-violating phenomena without denying locality. There \textit{is} another way out - the admitting of the relevance of history - which does not follow as an arguably \textit{ad-hoc} response to a problem, but as a consequence of our relatively more careful and principled formulation of the principle of Probabilistic Locality. 

Some commentators have maintained that the lesson of Bell-violating experiments is that some form of ``realism'' is untenable.\footnote{One clear exposition of such a view is found in Werner (2014), to which Maudlin (2014) gave a critical reply. For another detailed criticism of such views, see Norsen (2007).} The thought is that ``realism'' figures among the assumptions of Bell's theorem, and that by dropping this assumption, no problematic conclusion follows: the very task of finding out the localized events or local beables that figure in Bell-violating phenomena presumably betrays the ``realist'' presupposition that there are such events or beables to be found in the first place. This however seems to rest on a confusion. Particular instances of Bell-violating phenomena are as real as my current particular writing, and both of these, in virtue of being spatiotemporally extended, allow for an analysis into relatively well localized event-occurrences that constitute them. While the event-occurrences thereby found might radically defy our expectations - they might for instance not be describable in terms of objects bearing certain properties somewhere and somewhen - their reality cannot be denied, if the reality of the original spatiotemporally extended phenomenon is to be maintained, as it ought to be.

Passing from dissimilarities to similarities with the existing literature, our proposal can be viewed as relatively well aligned with a series of works that share our aim of arguing that Bell's original criterion of local causality is not the right probabilistic explication of what relativity teaches us in the first place. Nevertheless, the alternative principles so far proposed differ substantially from ours. Some of them explicitly refer to agent-centered notions, such as ``controllability'': while these might as well amount to legitimate and possibly valid principles, they are certainly not to replace the physical principle that Bell was after, and that we are after.\footnote{Jarrett (1984) and Shimony (1986) are classic references that suggested that Bell-inequality-violating phenomena may be compatible with relativity, as these need not exhibit ``uncontrollable'' non-locality. It is needless to say that Bell's classic (2004) is permeated by reasons to resist such suggestions. For more recent and developed agent-centered principles, see Friedrich (2015) and Healey (2017, p. 179-183).} Other formulations are articulated in terms of systems or objects, which, again, while possibly valid, fall short of our search for arguably more fundamental event-principles, principles that do not presuppose possibly problematic partitions into temporally subsisting systems-objects.\footnote{For example, a systems-centered principle of locality - one that articulates locality in terms of ``causal influences'' among ``systems'' - has recently been proposed by Barandes (2024).} 

Another similarity that our proposal bears to previous ones is our push for the acceptance that history is relevant, which appears to resemble previous discussions on the ``non-Markovianity'', the ``temporal non-locality'' and the ``indivisibility'' of quantum phenomena, as well as a recent suggestion that Bell-correlations are to be understood as ``selection artefacts''.\footnote{A relatively early suggestion that ``non-Markovianity'' might play an important role in understanding Bell-inequality-violating phenomena can be found in Suppes (1993). Adlam (2018) argued that embracing ``temporal non-locality'' offers a chance not only for an adequate understanding of quantum phenomena, but also for the development of novel physical theories. Barandes (2025) recently proposed that quantum phenomena are best understood as ``indivisible'' stochastic processes, processes that cannot be analyzed into a concatenation of sub-processes, leading to an essential dependence on history (for a critical reflection on this proposal, see Albert, 2026). Finally, Huw Price (2026)'s recent proposal that Bell-correlations can be viewed as selection artefacts appears to closely resemble our proposal that history is relevant, in so far as the actual course of history can be understood as what does the said selection, and what thereby induces Bell-correlations.} A due comparative analysis is however out of scope here, especially as these proposals would need to first be translated into our framework of probabilistic world-theories before allowing for an adequate comparison. It is only to be cautionarily noted that our work does not only assert the relevance of history, but also combines it with - and actually derives it from - a form of its \textit{ir}relevance, as manifested in the principle of Probabilistic Locality: the history of events that actually happened \textit{is} relevant, but the history of their probability distributions is \textit{not}. Our proposal is thus more nuanced than one that might flatly assert that events somehow influence each other non-locally in time.

Let us finish with a few comments on ways in which our work is still to be developed. As mentioned in section II, our choice to formulate the framework of world-theories with respect to a fixed spatiotemporal background $\mathcal{M}$ was not of essential importance. Indeed, instead of defining the worlds posited by a theory in terms of configurations of a list of events spread across a fixed space and time, each world might instead be defined directly in terms of the events that occur in that world and in terms of the spatiotemporal relations that obtain among them. Such a relational formulation of probabilistic world-theories would be more adequate, in that it would not posit an arguably superfluous entity - a spatiotemporal background - and would also more naturally allow for spatiotemporal relations to be dynamical, taking in the lessons of general relativity.

%One ought to furthermore inspect how our proposal bears on other experiments that conflict with Bell's criterion of local causality, such as the GHZ-experiment

%Another avenue GHZ. . . 

Furthermore, while the details of our proposal are definitely still to be polished, \textit{if} its gist is correct, then it reveals that there might after all be room for the development of a theory of quantum phenomena that speaks of localized event-occurrences, without explicit reference to instrumental notions such as preparations and measurements, and without conflict with relativity theory: there might, in other words, be room for something like a ``local-realist'' theory of the quantum, provided that the ``local'' is correctly formulated.\footnote{The ``might'' above is a reminder that a more systematic and encompassing investigation is needed before a stronger claim can be made. One ought to, for instance, study how our criterion of Probabilistic Locality squares with \textit{other} phenomena that appear to be in tension with locality, such as the GHZ-experiment. One ought to, as well, study whether there are some \textit{other} seemingly plausible assumptions, besides the Irrelevance of History, that might collapse our criterion into Bell's old criterion, and whose denial might not be as intelligible.} Our work is accordingly to be understood as an invitation to develop some such theory, to develop a local and potentially time-symmetric probabilistic world-theory that can account for quantum phenomena. Of course, before doing so, one ought to also evaluate how existing formulations of quantum theory fit into our framework. While many of the existing accounts-interpretations-formulations are currently insufficiently developed to allow for such an evaluation - relational, many-worlds and QBist accounts arguably do not determine in a precise way the events that possibly take place in any one quantum phenomenon - \textit{some} extensions-modifications of quantum mechanics nevertheless \textit{do} seem to be sufficiently mature, such as Bohmian and spontaneous-collapse theories. A due analysis thereof is however left for another occasion.\footnote{While another occasion is indeed needed for an extensive analysis, it does seem that no Bohmian theory can satisfy our criterion of Probabilistic Locality, simply due to the scarcity of the events that such theories posit. Take for example an experiment in which a single particle - at some time $t_1$ - is in a superposition state between two locations, and is let - at some later time $t_2$ - to interfere on a beam-splitter. The probability that the particle is detected at one particular port of the beam-splitter at $t_2$ depends on the \textit{phase-difference} between the two branches of the particle's wave-function at $t_1$. This phase-difference is however not in any way reflected in the distribution of events at $t_1$, since the only events posited by Bohmian theories amounts to the density of particles present somewhere-somewhen. Therefore, the distribution of events at $t_1$ does not at all determine the distribution of events at $t_2$, let alone in a way that would be compatible with relativistic causal structure. Bohmian theories accordingly appear to conflict with Probabilistic Locality, and they do so already in their account of the just sketched simple experiment that features no entanglement.}

\newpage

\section*{Acknowledgments}
This research was funded in whole or in part by the Austrian Science Fund (FWF) [10.55776/DOC162]. For open access purposes, the author has applied a CC BY public copyright license to any authoraccepted manuscript version arising from this submission.

\section*{References}

Adlam, E. (2018). Spooky action at a temporal distance. \textit{Entropy, 20}(1), 41.

\smallskip

Albert, D. Z. (2026). Quantum mechanics without a wave-function. \textit{Synthese, 208}(3), 141.

\smallskip

Baas, A., \& Le Bihan, B. (2023). What does the world look like according to superdeterminism?. The \textit{British Journal for the Philosophy of Science, 74}(3), 555-572.

\smallskip

Barandes, J. A. (2024). New prospects for a causally local formulation of quantum theory. \textit{arXiv preprint arXiv:2402.16935.}

\smallskip

Barandes, J. A. (2025). The Stochastic-Quantum Correspondence. \textit{Philosophy of Physics, 3}(1), Article 8. 

\smallskip

Bell, J. S. (2004). \textit{Speakable and Unspeakable in Quantum Mechanics}, 2nd edition, Cambridge: Cambridge University Press.

\smallskip

Butterfield, J. (2007). Stochastic Einstein locality revisited. \textit{The British Journal for the Philosophy of Science, 58}(4), 805-867.

\smallskip

Callender, C. (2026). Thermodynamic Asymmetry in Time. \textit{The Stanford Encyclopedia of Philosophy} (Summer 2026 Edition), Edward N. Zalta \& Uri Nodelman (eds.), URL = <\url{https://plato.stanford.edu/archives/sum2026/entries/time-thermo/}>. 

\smallskip

Friederich, S. (2015). Re-thinking local causality. \textit{Synthese, 192}(1), 221-240.

\smallskip

Ghirardi, G. C., Rimini, A., \& Weber, T. (1980). A general argument against superluminal transmission through the quantum mechanical measurement process. \textit{Lettere al Nuovo Cimento (1971-1985), 27}(10), 293-298.

\smallskip

Halvorson, H., Manchak, J. B., \& Weatherall, J. O. (2026). Deterministic theories. \textit{Philosophy and Phenomenological Research, 113}(1), 134-152.

\smallskip

Healey, R. (2017). \textit{The quantum revolution in philosophy}. Oxford University Press.

\smallskip

Hellman, G. (1982). Stochastic Einstein-locality and the Bell theorems. \textit{Synthese, 53}(3), 461-503.

\smallskip

Henson, J. (2005). Comparing causality principles. \textit{Studies In History and Philosophy of Science Part B: Studies In History and Philosophy of Modern Physics}, 36(3), 519-543.

\smallskip

Jarrett, J. P. (1984). On the physical significance of the locality conditions in the Bell arguments. \textit{Noûs, 18}(4), 569-589.

\smallskip

Maudlin, T. (2011). \textit{Quantum non-locality and relativity: Metaphysical intimations of modern physics}. John Wiley \& Sons.

\smallskip

Maudlin, T. (2014). Reply to comment on ‘What Bell did’. \textit{Journal of Physics A: Mathematical and Theoretical, 47}(42), 424012.

\smallskip

Myrwold, W., Genovese, M., Shimony, A. (2024). Bell’s Theorem. \textit{The Stanford Encyclopedia of Philosophy} (Summer 2024 Edition), Edward N. Zalta \& Uri Nodelman (eds.), URL = <\url{https://plato.stanford.edu/archives/sum2024/entries/bell-theorem/}>.

\smallskip

Norsen, T. (2007). Against ‘realism’. \textit{Foundations of Physics, 37}(3), 311-340.

\smallskip

Norsen, T. (2011). John S. Bell’s concept of local causality. \textit{American Journal of Physics, 79}(12), 1261-1275.

\smallskip

Price, H. (1996). \textit{Time's arrow \& Archimedes' point: new directions for the physics of time}. Oxford University Press.

\smallskip

Price, H. (2026). Retrieving the Baby: Reichenbach's Principle, Bell Locality, and Selection Bias. \textit{arXiv preprint arXiv:2602.16985.}

\smallskip

Shimony, A. (1984). Controllable and uncontrollable non-locality. In \textit{Foundations of Quantum Mechanics in the Light of New Technology}, S. Kamefuchi (ed.), Tokyo, The Physical Society of Japan, 225–230.

\smallskip

Suppes, P. (1993). Probabilistic causality in quantum mechanics. In \textit{Models and Methods in the Philosophy of Science: Selected Essays} (pp. 327-337). Dordrecht: Springer Netherlands.

\smallskip

Waegell, M., \& McQueen, K. J. (2025). From statistical dependence to the space of possible superdeterministic theories. \textit{European Journal for Philosophy of Science, 15}(4), 56.

\smallskip

Werner, R. F. (2014). Comment on ‘What Bell did’. \textit{Journal of Physics A: Mathematical and Theoretical, 47}(42), 424011.

\end{document}